%% file: o1-cbc-dq-paper.tex
\documentclass{iopart}

\expandafter\let\csname equation*\endcsname\relax
\expandafter\let\csname endequation*\endcsname\relax

\usepackage[acronym,shortcuts,sanitize=none]{glossaries}
\usepackage[colorlinks,linkcolor={blue},citecolor={blue},urlcolor={red}]{hyperref}
\usepackage{iopams}
\usepackage{subfig}
\usepackage[margin=10pt,font=small,format=hang]{caption}
\usepackage{multicol}
\usepackage{multirow}
\usepackage{array}
\usepackage{graphicx}
\usepackage[dvipsnames]{xcolor}
\usepackage{cleveref}
\usepackage{lineno}
\usepackage{hhline}
\pdfoutput=1

\DeclareGraphicsExtensions{%
    .pdf,.PDF,%
    .png,.PNG,%
    .jpg,.mps,.jpeg,.jbig2,.jb2,.JPG,.JPEG,.JBIG2,.JB2}

\begin{document}

\pagenumbering{arabic}

\title[Data Quality for CBC Searches in O1]{Effects of Data Quality Vetoes on a Search for Compact Binary Coalescences in Advanced LIGO's First Observing Run}
\input{LSC_Feb2016_Virgo_Feb2016-cqg.tex}

\begin{abstract}

The first observing run of Advanced LIGO spanned 4 months, from September 12, 2015 to January 19, 2016,  
during which gravitational waves were directly detected from two binary black hole systems,  
namely GW150914 and GW151226. Confident detection of gravitational waves requires an 
understanding of instrumental transients and artifacts that can reduce the sensitivity  
of a search. Studies of the quality of the detector data yield 
insights into the cause of instrumental artifacts and data quality vetoes specific to 
a search are produced to mitigate the effects of problematic data. In this paper, 
the systematic removal of noisy data from analysis time is shown 
to improve the sensitivity of searches for compact binary coalescences. 
The output of the 
PyCBC pipeline, which is a python-based code package used to search for gravitational 
wave signals from compact binary coalescences, is used as a metric for improvement. 
GW150914 was a loud enough signal that 
removing noisy data did not improve its significance. However, 
the removal of data with excess noise decreased the false alarm rate of GW151226 
by more than two orders of magnitude, from 1 in 770 years to less than 
1 in 186000 years.

\end{abstract}

\maketitle

%%%  ================= Introduction
%%% 
\section{Introduction}\label{sec:intro}

The Advanced Laser Interferometer Gravitational-Wave Observatory (aLIGO) is 
comprised of two dual-recycled 
Michelson interferometers \cite{TheLIGOScientific:2014jea} located in Livingston, LA (L1) and Hanford, WA (H1). 
A gravitational wave passing through a LIGO interferometer will induce a strain on spacetime, 
stretching and squeezing the 4~km arms and generating an interferometric signal at the 
antisymmetric port of the beamsplitter.  

Advanced LIGO's first observing run (O1) lasted from September 12, 2015 to January 19, 2016. 
A primary goal of this observing run was 
the detection of gravitational waves from compact binary coalescences (CBC) \cite{Babak:2012zx}. 
This goal was achieved with the detections of GW150914 and GW151226, both signals from 
binary black hole systems, which marked the 
first direct detections of gravitational waves \cite{GW150914-DETECTION,GW151226}. 
These detections were part of a 
broader search for CBC signals carried out by multiple search pipelines during O1 
\cite{Usman:2015kfa,Cannon2011Early,Privitera:2013xza,gstlal-methods,pycbc-software,Canton:2014ena} 
and searches for unmodeled transients \cite{GW150914-BURST,Klimenko:2008fu,Lynch:2015,BayesWave}.

Searching for gravitational waves requires an understanding of instrumental
features and artifacts that can adversely affect the output of a gravitational wave 
search pipeline. Throughout the observing
run, noisy data were identified in the form of data quality (DQ) vetoes  
to ensure that the analysis pipelines did not analyze data known to be 
contaminated with excess noise \cite{GW150914-DETCHAR}. 
These vetoes are discussed further in Section \ref{sec:vetoes}.
This study measures the effects of removing data with excess noise on the output 
of PyCBC \cite{pycbc-software, Usman:2015kfa, Canton:2014ena}, 
a python-based pipeline used to search for CBC signals. 
Section \ref{sec:pycbc} contains 
a brief description of the PyCBC search pipeline and its internal DQ features.

Section \ref{sec:selection} outlines the data selection and noise characterization 
processes. The DQ vetoes that are generated in the noise characterization 
process are described in Section \ref{sec:vetoes}.
The methodology of this study is discussed in Section \ref{sec:dq-effects}.
This paper focuses on two specific subsets of the O1 data set. The first data set, 
from September 12 - October 20, 2015, was used for background estimation 
for GW150914. This data set is discussed in Section \ref{sec:GW150914analysis}. The second 
data set, from December 3, 2015 - January 19, 2016, was used for 
background estimation for GW151226. This data set is discussed in Section \ref{sec:GW151226analysis}.
Section \ref{sec:limiting} describes the limiting noise sources for 
CBC searches.

\section{Data Selection}\label{sec:selection}

The strain data measured at the output of the detectors 
are typically non-stationary and 
non-Gaussian and contain noise artifacts of varying durations.
The longer duration non-stationary data can affect the overall sensitivity 
of the search, but they do not result in loud background events 
as they occur on a timescale of hours.
The transient noise artifacts, however, occur on a timescale of seconds and 
can reduce the sensitivity of 
CBC searches by producing loud background events. 

Data quality studies must be performed to search for causes of transients in the data that 
generate loud events in a gravitational wave search. If the source of 
noise is identified, a veto is generated to flag times when transient noise makes 
the data unsuitable for analysis. 
Section \ref{sec:vetoes} describes DQ vetoes that
are used to indicate when the detector data are known to have excess noise 
\cite{Nuttall:2015dqa,S6DetChar,GW150914-DETCHAR,Amaldi}.
The exception to this process is gating \cite{Usman:2015kfa}, which is a feature 
internal to the CBC searches. This gating, which is applied independently of DQ vetoes, 
uses a window function to remove times containing large transients from the input data stream.

\section{The PyCBC search pipeline}\label{sec:pycbc}

The PyCBC pipeline is designed to search for gravitational wave transients from CBCs \cite{Usman:2015kfa}. 
It employs a matched filter algorithm, which correlates expected CBC 
waveforms with detector data and outputs a ranking statistic, the signal-to-noise ratio (SNR). 
If the ranking statistic exceeds a specified threshold, an event, or ``trigger", is generated. 
The SNR of each trigger is weighted based on a signal consistency test \cite{Allen:2004gu}, 
resulting in a refined ranking statistic called re-weighted SNR. 
Section \ref{sec:chisq} discusses this signal consistency test further.

To perform this search, the matched filter algorithm needs to know what to search for. 
A collection of model CBC waveforms is generated before the analysis \cite{Taracchini:2013,Purrer:2015tud}. 
Each of these waveforms is called a template and 
the full collection of waveforms is referred to as the template bank. This template bank 
is constructed to span the astrophysical parameter space included in the search 
\cite{GW150914-CBC}. Each waveform is defined by the mass and spin of each compact 
object in the binary system. It is often convenient to combine the effects of each 
object's spin into one parameter called effective spin $\chi_{\mathrm{eff}}$, 
which is the mass-weighted spin of the system \cite{Privitera:2013xza}.  
The mass of the binary system is often represented by the chirp mass 
$\mathcal{M}$ \cite{PhysRev.131.435}, which is used to 
parameterize gravitational wave signals in general relativity. 

The search algorithm is run separately at each detector and a set of single detector  
triggers is generated. The two sets of single detector triggers are then compared to 
search for any events that were recorded within a 15 ms coincidence window, which 
reflects the travel time of a gravitational wave between the detectors and allows 
for uncertainty in the arrival time of a signal \cite{Usman:2015kfa}.
Any triggers that are found in coincidence with the same source parameters 
in both detectors represent potential gravitational wave signals 
and are referred to as foreground events. Some of these foreground events will be 
chance coincidences between noise in each detector, which is expected given 
the number of events in each data set. 

To determine the statistical significance of foreground events, 
a background distribution is generated using a time shift technique \cite{Usman:2015kfa}. 
The statistical significance of any candidate gravitational wave is then 
quantified by calculating the rate of background events from detector noise that 
are at least as loud as the candidate event. 
This statistic is called the false alarm rate (FAR). 
Any loud triggers that appear as the result of instrumental transients will extend 
the background distribution and the influence the measured false alarm rate.
The purpose of the DQ effort as a whole is thus two-fold: to ensure that the 
search is using representative 
detector data in the background noise estimation and to suppress the rate of loud events 
that will pollute both the background and the foreground distributions. 

\subsection{$\chi^{2}$ signal consistency test}\label{sec:chisq}

A further layer of effective DQ that is internal to the PyCBC pipeline is the application of the 
$\chi^{2}$ signal consistency test \cite{Allen:2004gu}. The SNR produced by the matched filter is an 
integral in the frequency domain. The $\chi^{2}$ test divides each CBC waveform into 
frequency bins of equal power, checking that the SNR is distributed as a function of frequency 
as expected from an actual CBC signal. 
Each trigger that comes out of the matched filter search is down-weighted based on the results of 
the $\chi^{2}$ test. This is folded into a new ranking statistic for CBC triggers, 
which is called re-weighted SNR and is denoted by $\hat{\rho}$.
The ranking statistic for coincident events in the PyCBC search is the network 
re-weighted SNR, $\hat{\rho}_{c}$, which is the quadrature sum of the re-weighted 
SNR from each detector. Since a real signal has a power 
distribution that matches the template waveform, it will not be 
down-weighted by the $\chi^{2}$ test; the SNR and the re-weighted SNR will be the same.

\begin{figure}[!ht]%
\centering
  \subfloat[]{
      \includegraphics[width=.75\textwidth]{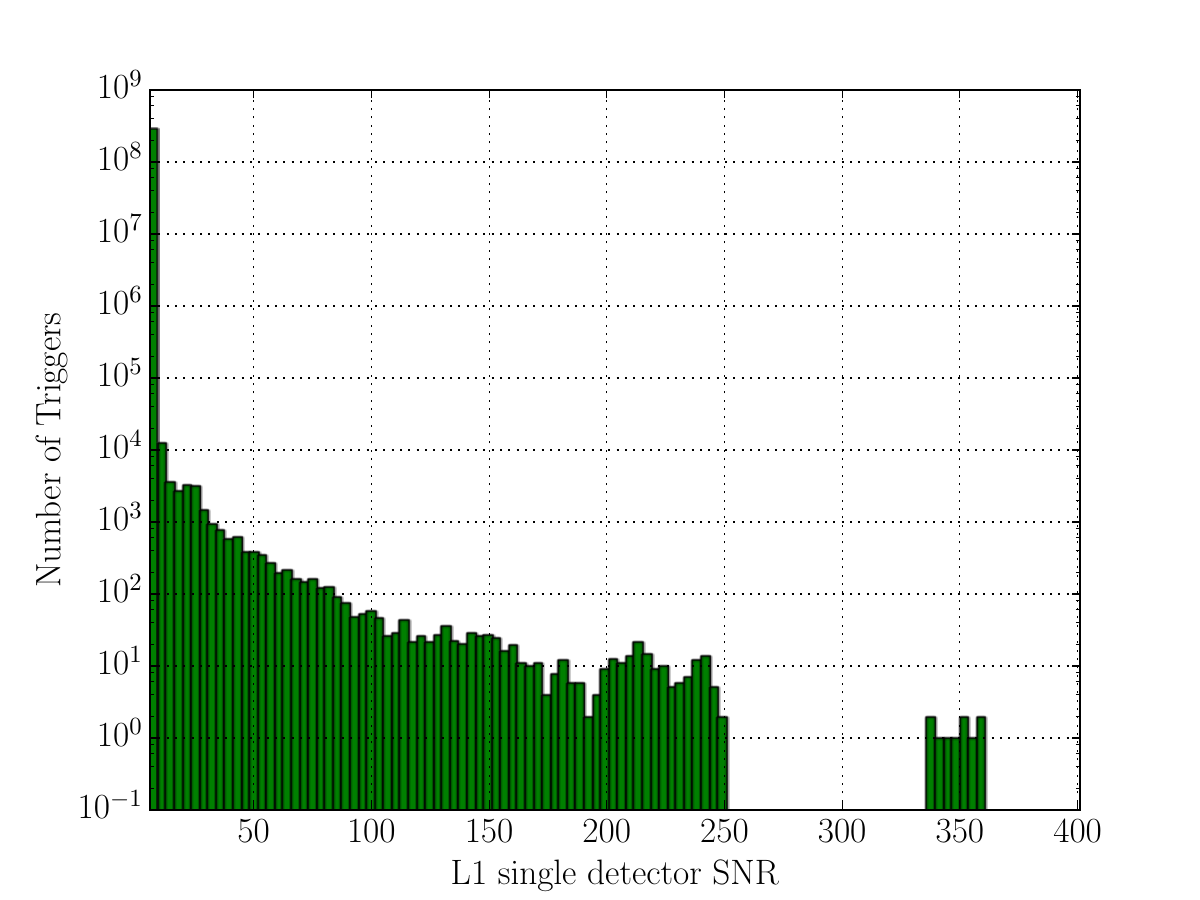}
      \label{subfig:l1-snr-hist}
  }
  
  \subfloat[]{
      \includegraphics[width=.75\textwidth]{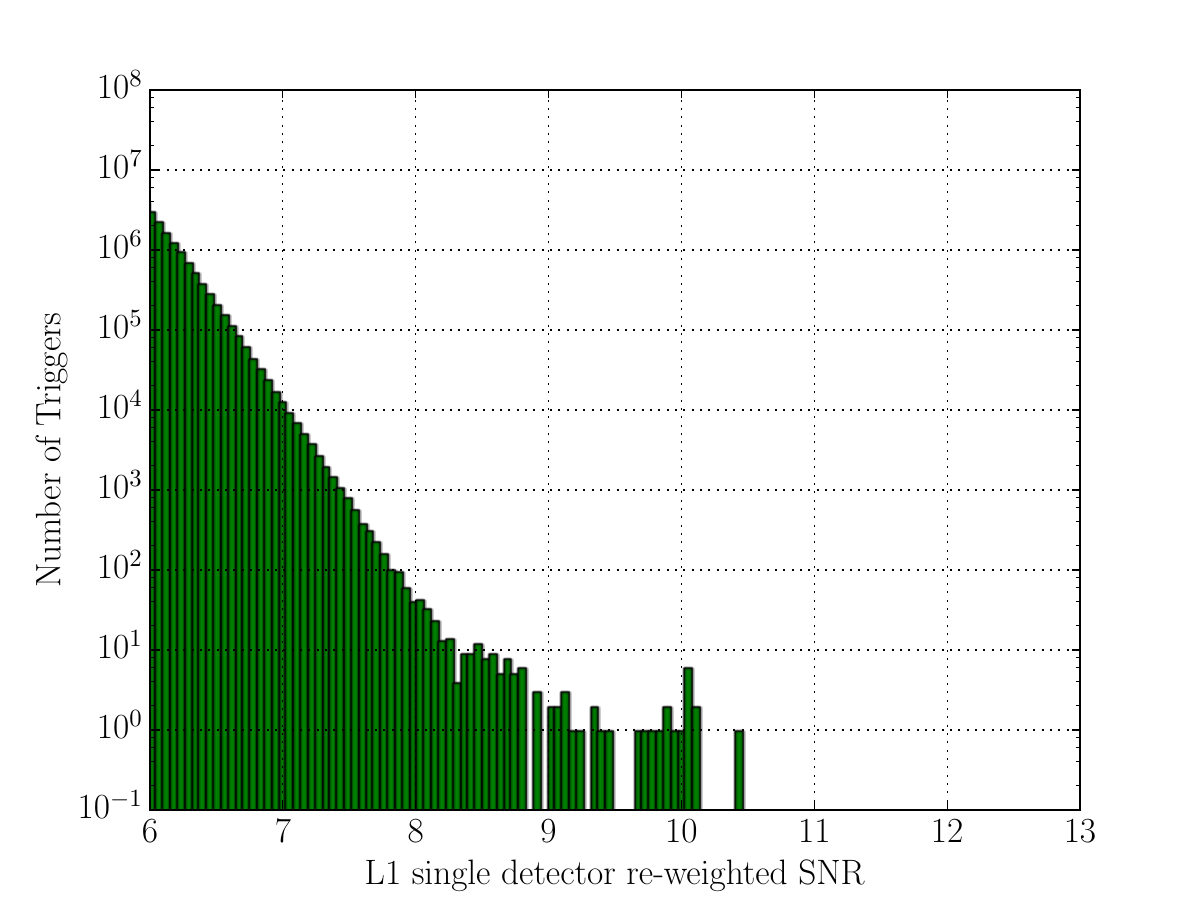}
      \label{subfig:l1-newsnr-hist}
  }
  
  \caption{Histograms of single detector triggers from the Livingston (L1) detector. %
           These triggers were generated using data from September 12 to October 20, 2015. These histograms %
           contain triggers from the entire template bank, but %
           exclude any triggers found in coincidence between the two detectors. %
           (\ref{subfig:l1-snr-hist}) A histogram of single detector triggers in SNR. %
           The tail of this distribution extends beyond SNR = 100. %
           (\ref{subfig:l1-newsnr-hist}) A histogram of single detector triggers in re-weighted SNR. %
           The chi-squared test down-weights the long tail of SNR triggers %
           in the re-weighted SNR distribution. The triggers found using only the %
           Hanford detector have a similar distribution.}
  \label{fig:cbc-newsnr-histograms}
\end{figure}

This test is extremely powerful, as shown in Figure \ref{fig:cbc-newsnr-histograms}, which shows the 
distribution of single detector PyCBC triggers generated from September 12 to October 20, 2015. 
Figure \ref{subfig:l1-snr-hist} shows the distribution of triggers in SNR. The extensive tail of 
triggers with high SNR, which extends beyond SNR 100, is down-weighted in the re-weighted SNR distribution, 
leaving behind a tail that extends to $\hat{\rho} \approx$ 10.5 as seen in Figure \ref{subfig:l1-newsnr-hist}. 
This re-weighted SNR tail represents the loudest single detector background triggers in the CBC search. 
Investigating this set of loudest background triggers guides DQ efforts in defining the current limiting 
noise sources to the CBC search.

\section{Data quality vetoes}\label{sec:vetoes}

As seen in Figure \ref{fig:cbc-newsnr-histograms}, the $\chi^{2}$ test is a powerful tool, but 
there is still a considerable tail in the single detector trigger distribution. 
This tail is often caused by transient instrumental noise. If these noise sources can be linked to a 
systematic instrumental cause or a period of highly irregular instrumental performance, 
they can be flagged and removed from the analysis in the form of a 
DQ veto.

DQ vetoes indicate times that are unsuitable for analysis or are likely to 
produce false alarm gravitational wave triggers. These vetoes are constructed by 
$\sim$200,000 witness sensors that continuously monitor LIGO detectors and their 
environment \cite{GW150914-DETCHAR}. 
Before a witness sensor is used to generate a DQ veto, its sensitivity to a genuine 
gravitational wave signal is assessed to ensure true signals are not unnecessarily 
removed from analysis. To test this, a gravitational wave signal is 
simulated by electromagnetically controlling the motion of the detector optics. 
If a witness sensor is observed to be sensitive to such an injected signal, 
it will be considered unsafe for generating DQ vetoes.

As an example, a veto was generated during O1 to mark times when an electronics fault 
caused amplitude fluctuations in the radio frequency (RF) sidebands used to sense and control LIGO's 
optical cavities. These amplitude fluctuations introduced noise into the feedback loops controlling 
the motion of the LIGO optics. This excess motion coupled into the output of the detector and 
created transient, broadband noise artifacts. 
A witness sensor monitoring the amplitude stabilization control signal 
for the RF sidebands was found to be correlated to the excess noise in the detector output. 
A threshold was established to indicate when the amplitude fluctuations were 
significant enough to couple into the detector output and any times where the fluctuation 
amplitude exceeded 
this threshold were marked as unfit for astrophysical analysis. 
The process of generating this veto is further detailed in Appendix A of \cite{GW150914-DETCHAR}. 

Other examples of noise sources that were vetoed in O1 are 
photodiode saturations, analog-to-digital converter (ADC) and digital-to-analog 
converter (DAC) overflows, elevated seismic noise, and computer failures. 
These DQ vetoes were generated using a process similar to that used to veto the RF 
amplitude fluctuation noise. 

DQ vetoes are produced for all analysis time based 
on systematic instrumental conditions without any regard for the presence of gravitational wave 
signals. All data are treated equally; the removal of data with excess noise has the ability to 
remove real gravitational wave signals as well as background events. 
The same set of DQ vetoes was applied by all CBC searches in O1 and will be released 
with any public data to ensure that astrophysical results are reproducible.  
Further details on DQ vetoes applied in the first observing run 
are available in a paper detailing the transient noise in the detectors at the 
time of GW150914 \cite{GW150914-DETCHAR}.

\section{Measuring the Effects of Data Quality Vetoes}\label{sec:dq-effects}

To test the effects of DQ vetoes, the PyCBC search pipeline was run 
with and without applying vetoes. The only vetoes that were used in all runs 
are those that indicate that the data were not properly 
calibrated, that a data dropout occurred, or that there were test signals being injected 
into the detectors. 
Gating is internal to the search pipeline and was applied in all of the analyses.
Two methods were used to understand the effects of applying vetoes. The first, 
described in Section \ref{sec:VT}, considers the average sensitivity of the search 
pipeline to gravitational wave signals. The second, described in Section 
\ref{sec:backgrounds}, compares the measured search backgrounds and the false alarm rates 
of recovered gravitational wave signals.

\subsection{Measuring search sensitivity}\label{sec:VT}

The metric used to measure the sensitivity of the search pipeline is 
sensitive volume. Sensitive volume is measured by 
injecting simulated gravitational wave signals into the data and attempting to 
recover them using the search \cite{Usman:2015kfa}. The ability of the 
pipeline to recover signals at a given false alarm rate is then measured by analyzing 
the number of missed and recovered injections.

In addition to the sensitive volume, the amount of time used in the analysis must 
be considered when removing noisy data. If a search is rejecting too much data, it 
will miss the opportunity to detect signals.  To address this, the sensitive 
volume of the search is multiplied by the amount of analysis time to create a new metric 
called VT. 
If time is removed from an analysis, 
the sensitive volume of the search must increase to make up for the shorter analyzed time.

The sensitivity of a search varies as a 
function of the significance threshold set for candidate events. The 
VT ratios are therefore calculated at both the 1 per 100 year and the 
1 per 1000 year levels. These significance levels are expressed as
inverse false alarm rates (IFAR). 

\subsection{Comparing search backgrounds}\label{sec:backgrounds}

In the first observing run,
the bank of CBC waveform templates used in the PyCBC search was divided into three bins 
\cite{GW150914-CBC}. The significance of any candidate gravitational wave found in 
coincidence between the two detectors is calculated relative to the background in its bin. 
Waveforms with different parameters will respond to instrumental transients in different ways. 
This binning is performed so that any foreground triggers are compared to a background 
generated from similar waveforms.
As such, the effects of removing data from the PyCBC search are variable depending on 
which bin is considered. The actual gravitational wave signals 
discovered in the PyCBC search,
GW150914 and GW151226, were part of a full search that was broken into 3 bins but reported
as a single table of results. Because of this, their reported false alarm rates include a 
trials factor of 3. The background distributions shown in Sections
\ref{sec:GW150914analysis} and \ref{sec:GW151226analysis} were measured on a bin-by-bin basis,
so the cumulative trigger rates have not been divided by 3. 

The first bin is called the binary neutron star (BNS) bin and contains all waveforms 
with $\mathcal{M} < 1.74$. 
The second bin is the edge bin, which is defined based on the peak frequency $f_{\mathrm{peak}}$ 
of each CBC waveform.
These waveforms are typically shorter in duration than binary neutron star waveforms 
and are comprised of
both binary black hole (BBH) and neutron star-black hole (NSBH) binary waveforms. 
Waveforms in the edge bin typically have high masses and negative $\chi_\mathrm{eff}$.
In this analysis, the edge bin contained waveforms with $f_{\mathrm{peak}} <$ 100 Hz. 
The third bin is the bulk bin, which contains all remaining 
waveforms needed to span the parameter space of the search. This contains BBH and NSBH 
waveforms with a variety of mass ratios and spins.

\section{Analysis containing GW150914}\label{sec:GW150914analysis}

The analysis containing GW150914 lasted from September 12 - October 20, 2015 and 
contained a total of 18.2 days of coincident detector data. Of this 18.2 days, 
7.7\% was considered unfit for astrophysical analysis and was removed from the 
data set. 

\subsection{Search sensitivity}

To measure the effects of DQ vetoes on the sensitivity of the search, 
the analysis containing GW150914 was performed with and without applying data 
quality vetoes. The resulting measurements of VT were divided to calculate a VT ratio. 

\begin{figure}[ht!]%
\centering
  \includegraphics[width=\textwidth]{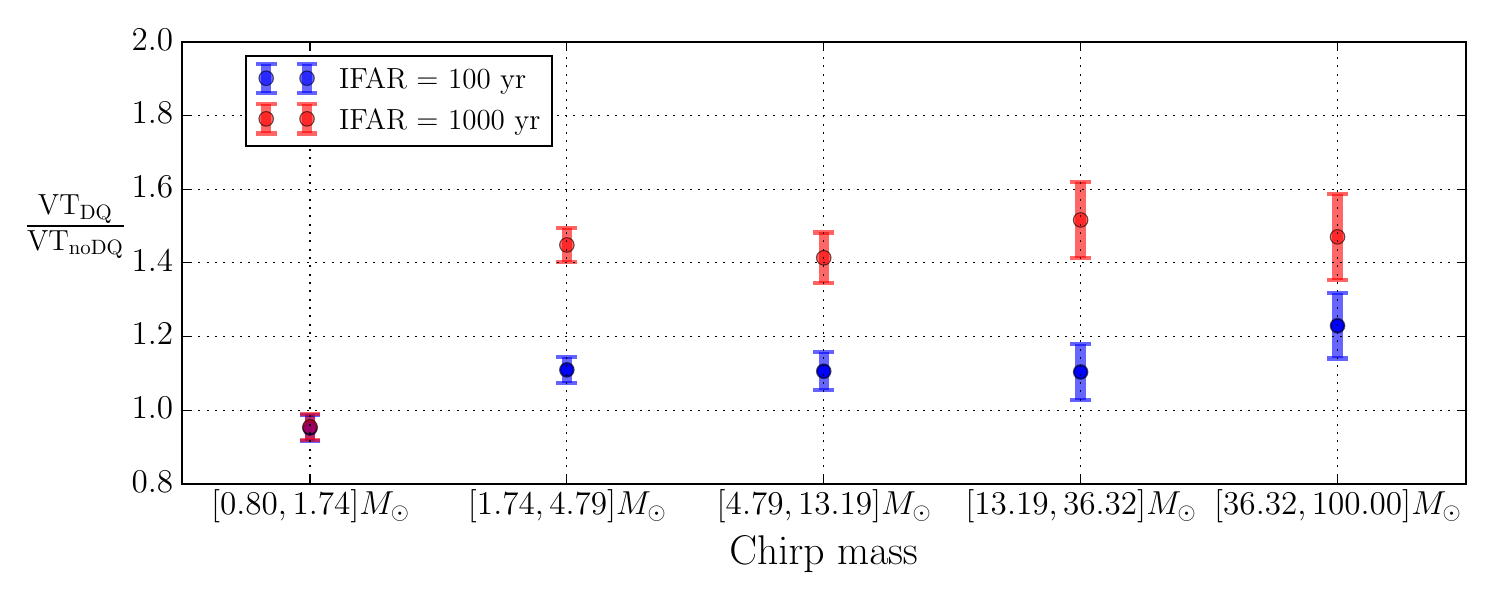}
  \caption{The change in search sensitivity when DQ vetoes are applied for 
           the analysis containing GW150914. The error bars show the 
           $1~\sigma$ error from each VT calculation combined in quadrature. 
           The lowest chirp mass bin, which contains BNS signals, does not show any 
           improvement in sensitivity. 
           For marginally significant signals at IFAR = 100, the measured value of 
           VT increases 
           by 3-32\% in higher chirp mass bins. For highly significant 
           signals at IFAR = 1000, the measured value of VT increases by 34-62\% 
           in higher chirp mass bins. 
          }
  \label{fig:GW150914-VT-RATIO}
\end{figure}

Figure \ref{fig:GW150914-VT-RATIO} shows the change in VT when vetoes are 
applied for two values of IFAR and several chirp mass bins. The lowest chirp mass 
bin contains BNS signals and does not show any improvement in sensitivity when 
DQ vetoes are applied. This is discussed further in section \ref{GW150914-BNS}.
Since the bulk and edge bins contain systems with a large range of masses that can respond
in different ways to instrumental artifacts, these higher mass bins have been split for 
sensitivity estimation. The higher mass bins in Figure \ref{fig:GW150914-VT-RATIO} are 
binned linearly in ln($\mathcal{M}$). The higher chirp mass bins show an improvement 
in search sensitivity for both values of IFAR. 

\subsection{BNS bin}\label{GW150914-BNS}

Binary neutron star systems have the longest waveforms used in the search pipelines. 
Since these signals spend $\sim 10 - 100$ seconds in LIGO's sensitive band, 
the $\chi^{2}$ test is effective at discriminating between binary neutron star signals 
and transient noise, which have a duration of $\sim 1$ second. 

\begin{figure}[!ht]%
\centering
  \subfloat[]{
  \includegraphics[width=0.495\textwidth]{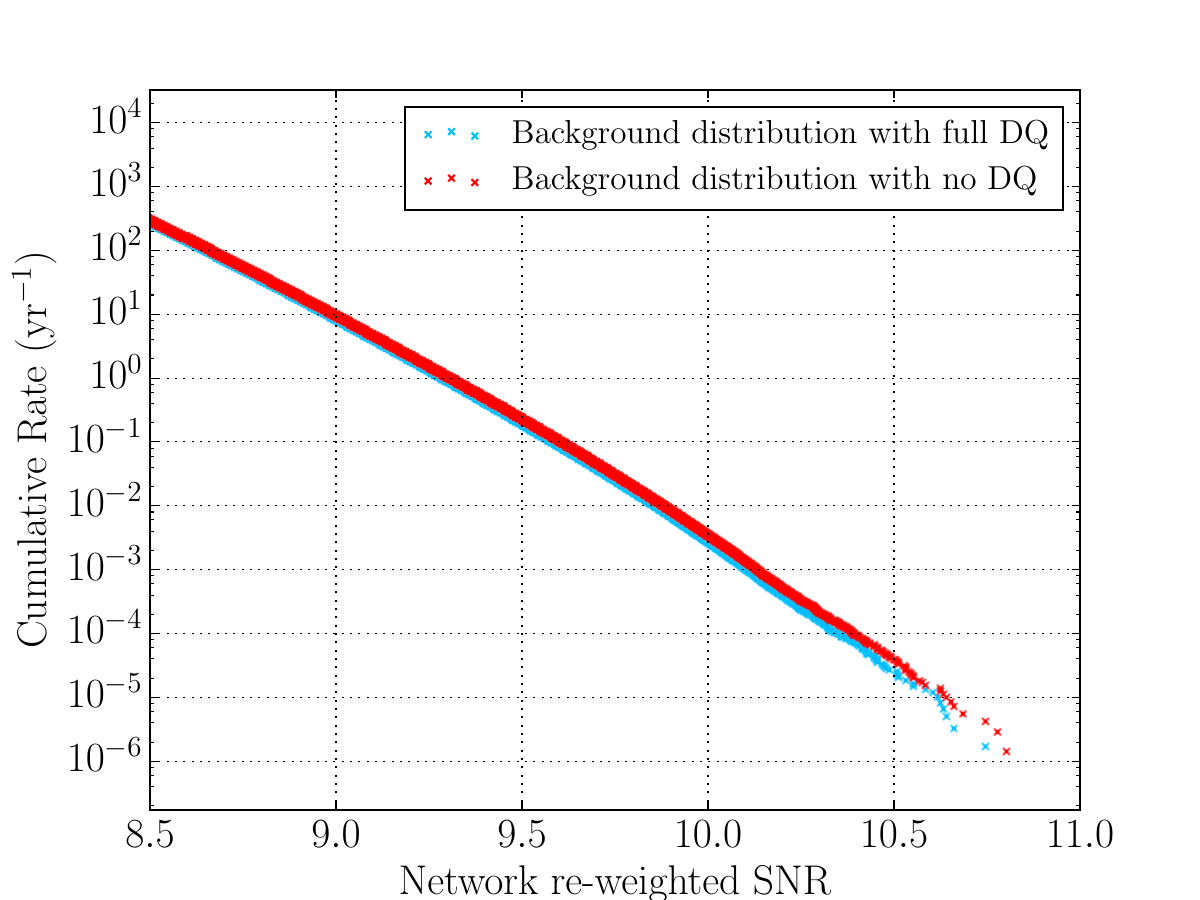}
  \label{subfig:bns-bin-gw150914-rate}
  }
  \subfloat[]{
  \includegraphics[width=0.495\textwidth]{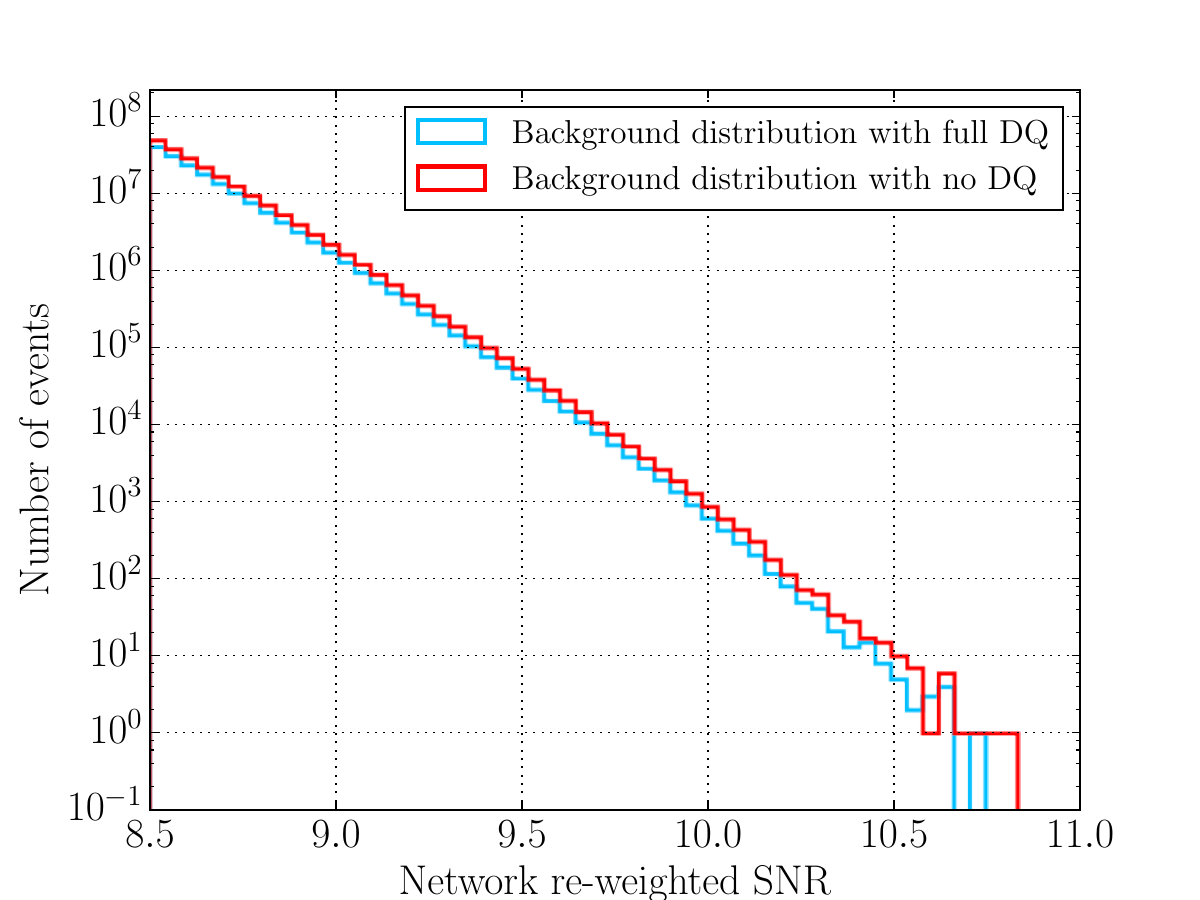}
  \label{subfig:bns-bin-gw150914-raw}
  }
  \caption{The background distribution in the BNS bin before and after applying DQ vetoes %
           for the analysis containing GW150914. %
           (\ref{subfig:bns-bin-gw150914-rate}) The cumulative rate of background triggers %
           in the BNS bin as a function of re-weighted SNR. %
           (\ref{subfig:bns-bin-gw150914-raw}) A histogram of background triggers %
           in the BNS bin. % 
           The red traces indicate the %
           distribution of background triggers without noisy data removed, %
           the cyan traces indicate the distribution of background triggers with %
           all DQ vetoes applied. The BNS bin shows %
           no significant improvement in cumulative rate.} 
  \label{fig:bns-bin-far_GW150914}
\end{figure}

Figure \ref{fig:bns-bin-far_GW150914} shows the background distribution of the BNS bin in the 
PyCBC search for the analysis containing GW150914. As expected, since waveforms in the BNS bin 
are not as susceptible to instrumental transients, the background distribution does not change 
significantly when noisy data are removed from the analysis.

\subsection{Bulk bin}\label{sec:bulk-bin}

\begin{figure}[!ht]%
\centering
  \subfloat[]{
  \includegraphics[width=0.495\textwidth]{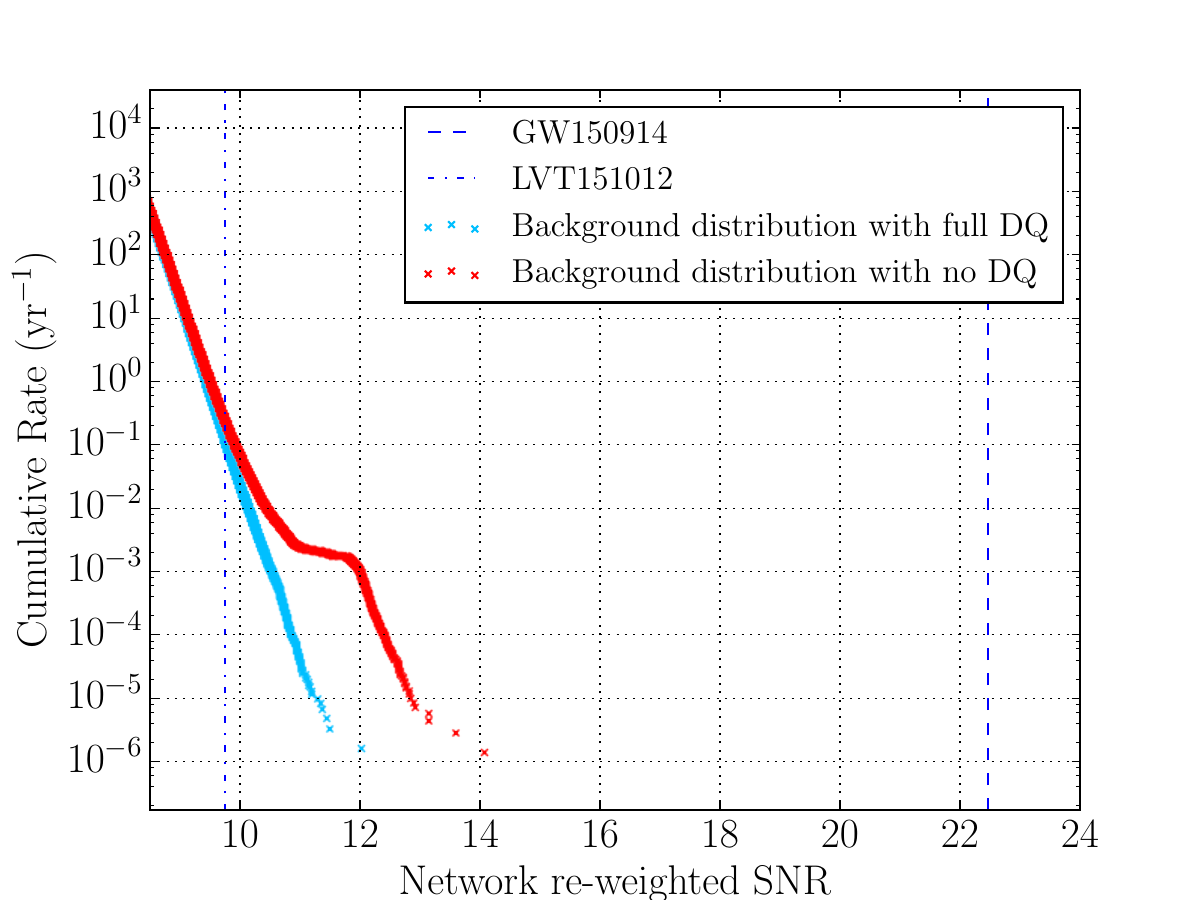}
  \label{subfig:bulk-bin-gw150914-rate}
  }
  \subfloat[]{
  \includegraphics[width=0.495\textwidth]{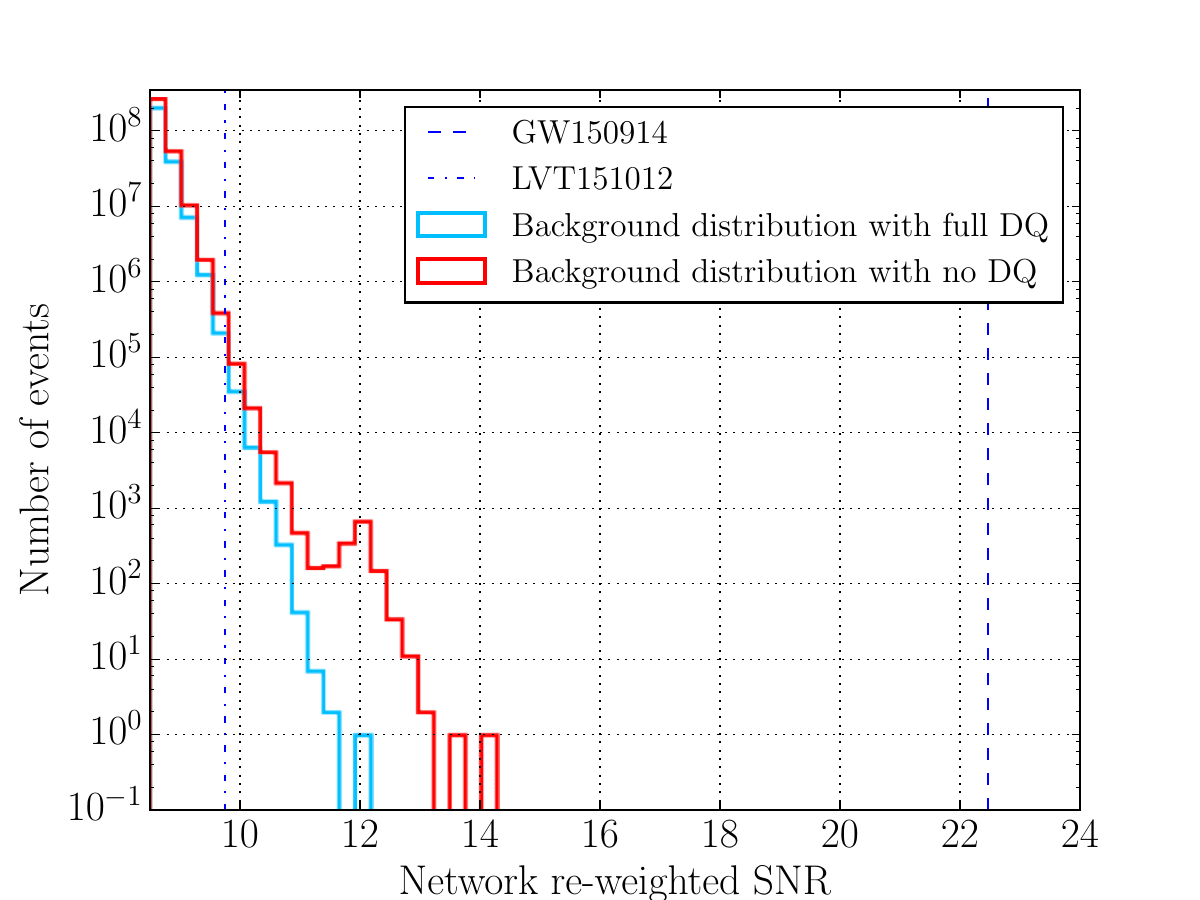}
  \label{subfig:bulk-bin-gw150914-raw}
  }
  \caption{The background distribution in the bulk bin before and after applying DQ vetoes %
           for the analysis containing GW150914. % 
           (\ref{subfig:bulk-bin-gw150914-rate}) The cumulative rate of background triggers %
           in the bulk bin as a function of re-weighted SNR. % 
           (\ref{subfig:bulk-bin-gw150914-raw}) A histogram of background triggers %
           in the bulk bin. %
           The red traces indicate the %
           distribution of background triggers without noisy data removed and the cyan traces %
           indicate the distribution of background triggers with all DQ vetoes applied. %
           When vetoes are not applied, there is a shoulder in the distribution %
           that limits the sensitivity of the search. % 
           The dash-dotted line indicates the network re-weighted SNR of LVT151012. %
           The dashed line indicates the network re-weighted SNR of GW150914, which is the loudest %
           event in this bin for both configurations.}
  \label{fig:bulk-bin-far_GW150914}
\end{figure}

Figure \ref{fig:bulk-bin-far_GW150914} shows the background distribution in the bulk 
bin for the analysis containing GW150914. 
If noisy data are not removed from the analysis, there is a shoulder in the distribution that 
extends to $\hat{\rho}_{c} =$ 14, which limits the sensitivity of the search in the region 
where 11 $< \hat{\rho}_{c} <$ 14. 

\subsubsection{LVT151012}\label{sec:LVT151012}
The second most significant trigger in the analysis containing GW150914 was LVT151012, 
recorded on October 12, 2015 \cite{GW150914-CBC,O1BBH}. 
This trigger was recovered in the bulk bin with $\hat{\rho}_{c} =$ 9.75 and a 
false alarm rate of 0.33 $\mathrm{yr}^{-1}$. This is not significant enough 
to be claimed as a confident detection. 
The false alarm rate decreases by a factor of 2.1 when DQ vetoes are 
applied, as shown in Table \ref{table:151012-far}.

\begin{table}[!ht]%
  \begin{center}
    \begin{tabular}{lc}
      \hline
      Analysis configuration & False alarm rate ($\mathrm{yr}^{-1}$) \\ \hline
      All vetoes applied & 0.33 \\ 
      No vetoes applied & 0.69 \\
      \hline
    \end{tabular}
  \end{center}
  \caption{Table of bulk bin false alarm rates for LVT151012. %
           } 
  \label{table:151012-far}
\end{table}

\subsection{Edge bin}\label{sec:edge-bin}

\begin{figure}[!ht]%
\centering
  \subfloat[]{
  \includegraphics[width=0.495\textwidth]{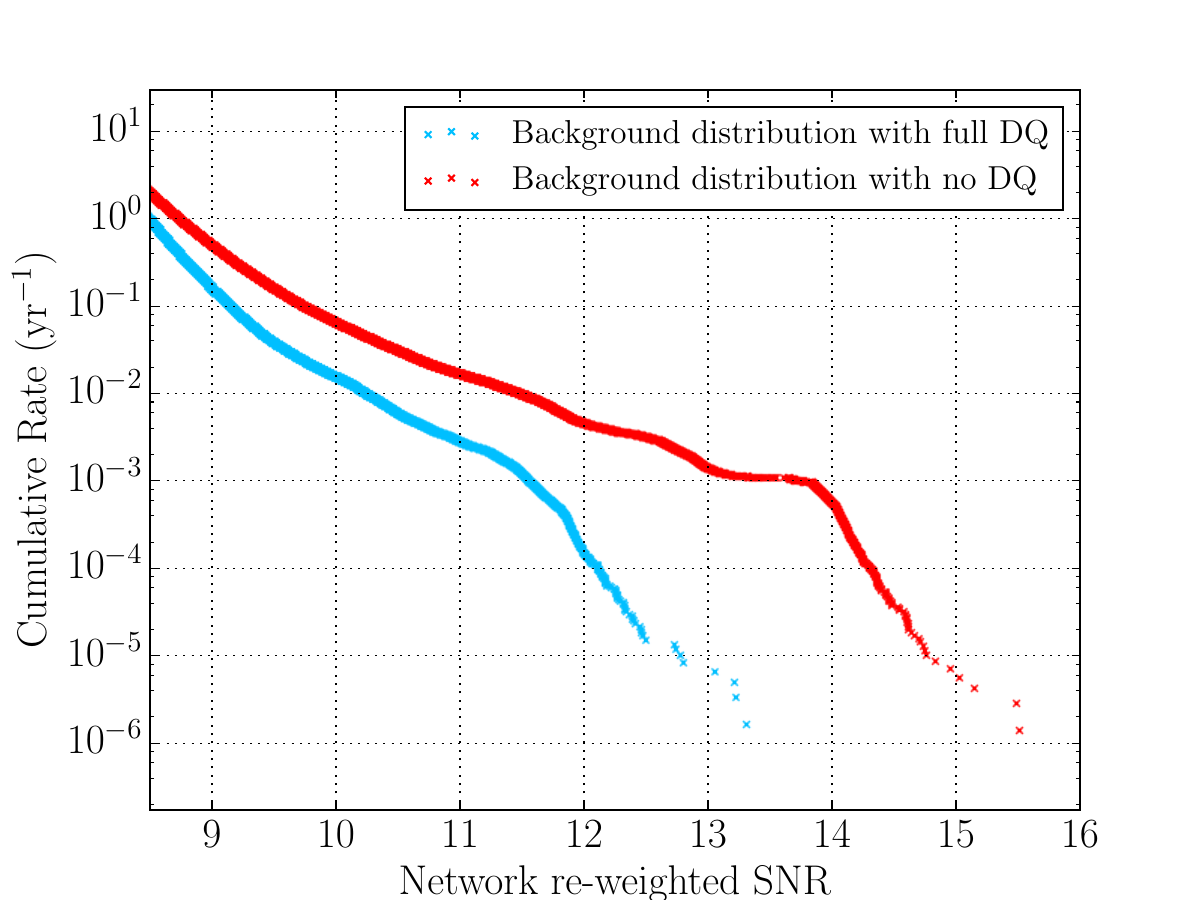}
  \label{subfig:edge-bin-gw150914-rate}
  }
  \subfloat[]{
  \includegraphics[width=0.495\textwidth]{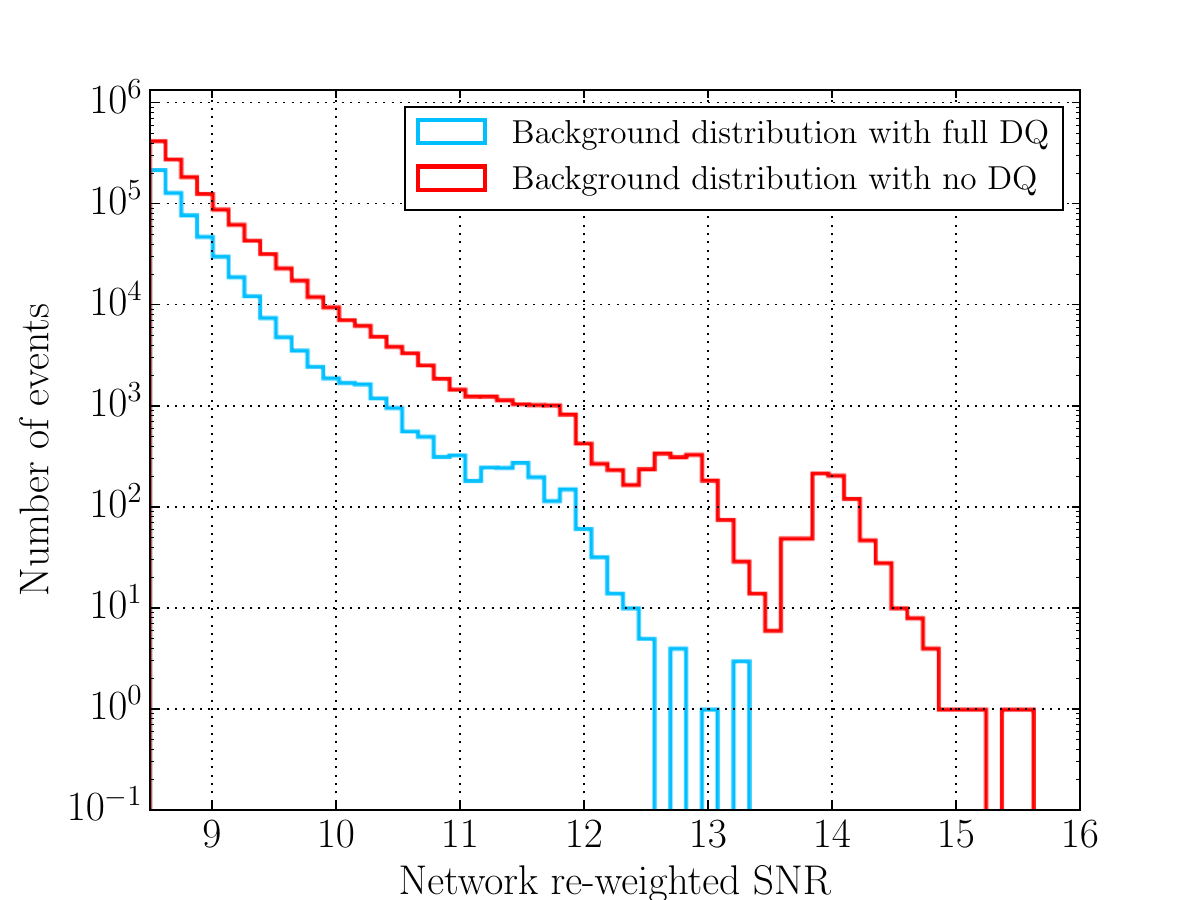}
  \label{subfig:edge-bin-gw150914-raw}
  }
  \caption{The background distribution in the edge bin before and after applying DQ vetoes %
           for the analysis containing GW150914. %
           (\ref{subfig:edge-bin-gw150914-rate}) The cumulative rate of background triggers %
           in the edge bin as a function of re-weighted SNR. %
           (\ref{subfig:edge-bin-gw150914-raw}) A histogram of background triggers %
           in the edge bin. % 
           The red traces indicate the %
           distribution of background triggers without noisy data removed from the analysis and %
           the cyan traces indicate the distribution of background triggers with all data % 
           quality vetoes applied. %
          }
\label{fig:edge-bin-far_GW150914}
\end{figure}

Figure \ref{fig:edge-bin-far_GW150914} shows the background 
distribution in the edge bin for the analysis containing GW150914. 
There is a visible separation between the two curves that 
increases for larger values of $\hat{\rho}_{c}$, indicating that the ability of the search pipeline 
to make confident detections is diminished in this region.

\subsubsection{GW150914}

The gravitational wave signal GW150914 was detected on September 14, 2015 with 
$\hat{\rho}_{c} =$ 23.6 \cite{GW150914-DETECTION}.  
The false alarm rate of GW150914 does not change significantly when noisy data 
are removed from the analysis, which can be seen in Table \ref{table:150914-far}. 
This is an expected result as GW150914 is louder than the entire background distribution 
in the bulk bin.

\begin{table}[!ht]%
  \begin{center}
    \begin{tabular}{lc}
      \hline
      Analysis configuration & False alarm rate ($\mathrm{yr}^{-1}$) \\ \hline
      All vetoes applied & $<~5.17\times10^{-6}$ \\
      No vetoes applied & $<~4.43\times10^{-6}$ \\
      \hline
    \end{tabular}
  \end{center}
  \caption{Table of bulk bin false alarm rates for GW150914. %
           GW150914 is loud enough that its false alarm rate %
           does not change significantly when noisy data are removed from the analysis. % 
           Any change in false alarm rate is due to small changes in the total analysis %
           time after data removal. %
           }
  \label{table:150914-far}
\end{table}

\section{Analysis containing GW151226}\label{sec:GW151226analysis}

The extended analysis containing GW151226 lasted from December 3, 2015 - January 19, 2016 and 
contained a total of 16.7 days of coincident detector data. 
Of this 16.7 days, 6.5\% was considered unfit for astrophysical analysis and was removed 
from the data set.

\subsection{Search sensitivity}

\begin{figure}[ht!]%
\centering
  \includegraphics[width=\textwidth]{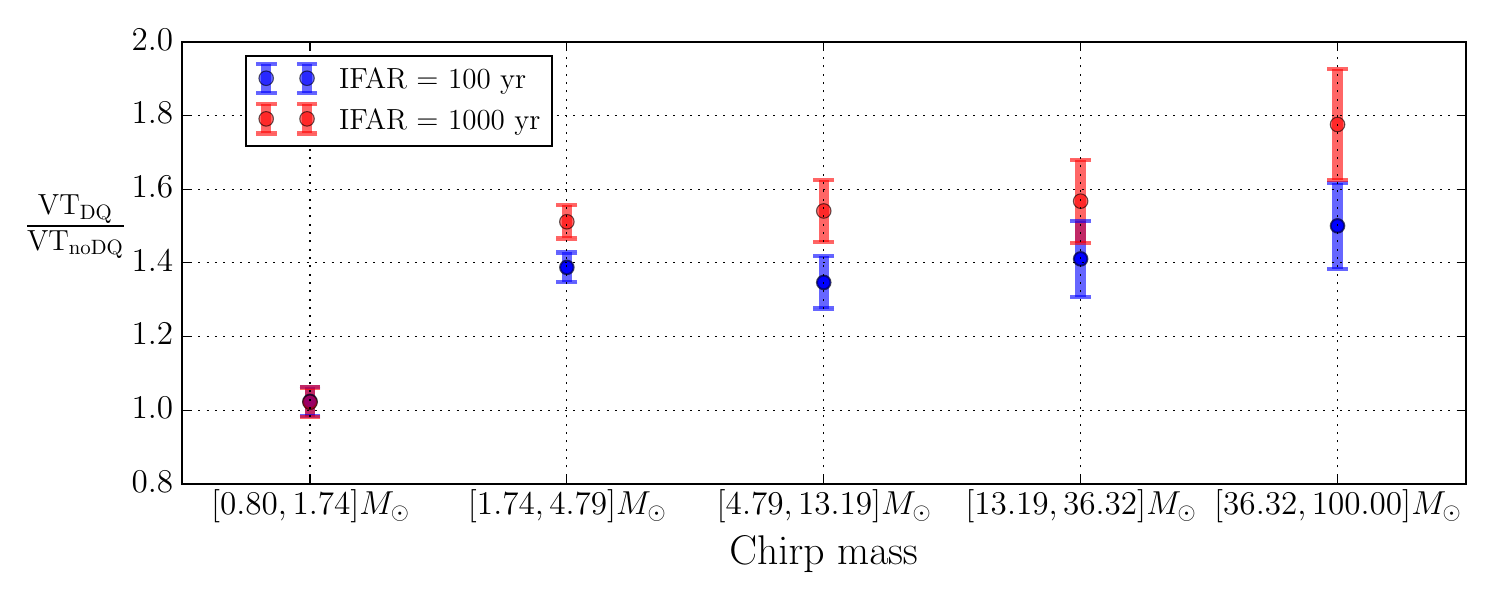}
  \caption{The change in search sensitivity when DQ vetoes are applied for
           the analysis containing GW151226. The error bars show the
           $1~\sigma$ error from each VT calculation combined in quadrature.
           The lowest chirp mass bin, which contains BNS signals, shows a small  
           improvement in sensitivity when vetoes are applied, 
           though the error bars are consistent with a VT ratio of 1. 
           For marginally significant signals at IFAR = 100, the measured value of
           VT increases
           by 27-62\% in higher chirp mass bins. For highly significant
           signals at IFAR = 1000, the measured value of VT increases by 45-90\%
           in higher chirp mass bins.
          }
  \label{fig:GW151226-VT-RATIO}
\end{figure}

Figure \ref{fig:GW151226-VT-RATIO} shows the change in VT when DQ vetoes are
applied to the analysis containing GW151226. For this analysis, the lowest chirp mass
bin, which contains BNS signals, shows a slight improvement when 
vetoes are applied. This improvement is discussed further in section 
\ref{GW151226-BNS}. Similar to the analysis containing GW150914, the higher chirp mass 
bins show an improvement in search sensitivity for both values of IFAR.

\subsection{BNS bin}\label{GW151226-BNS}

\begin{figure}[!ht]%
\centering
  \subfloat[]{
  \includegraphics[width=0.495\textwidth]{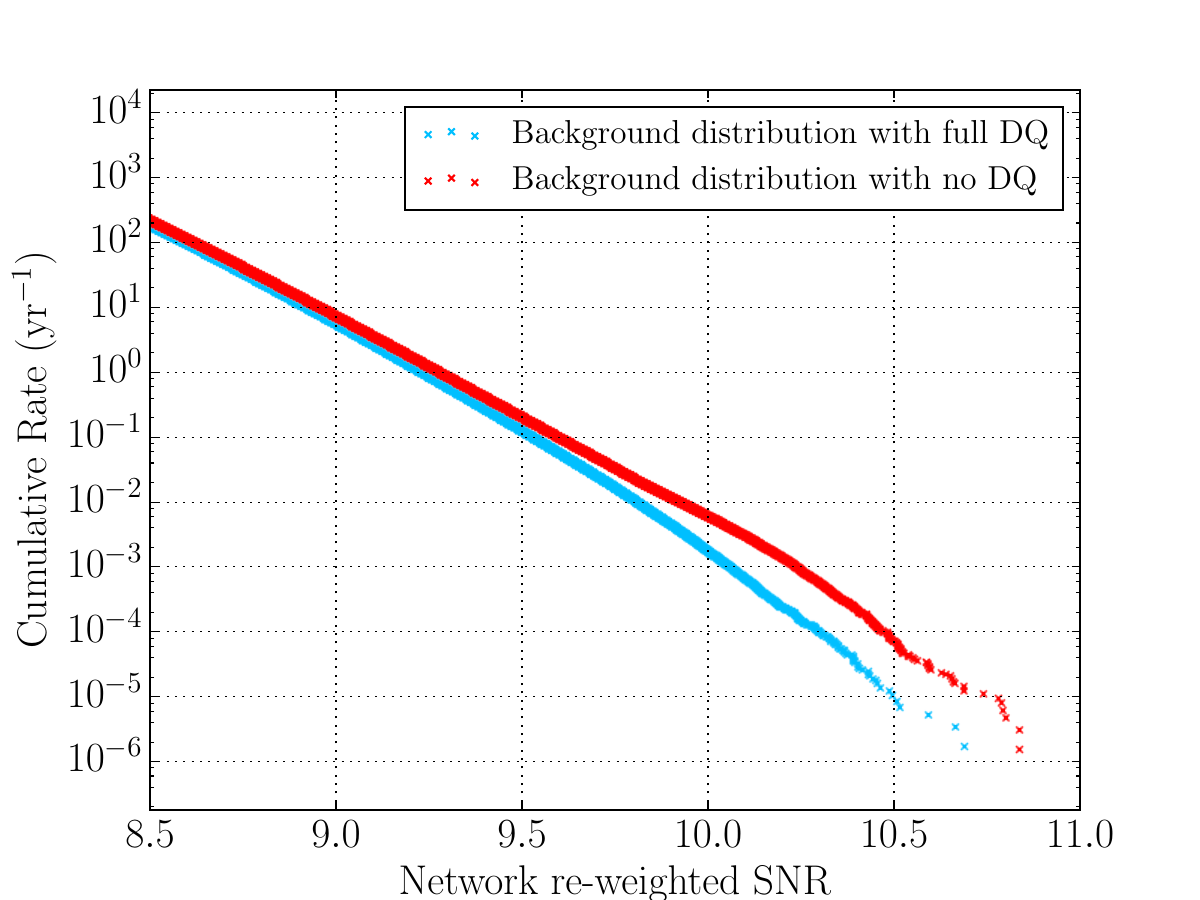}
  \label{subfig:bns-bin-gw151226-rate}
  }
  \subfloat[]{
  \includegraphics[width=0.495\textwidth]{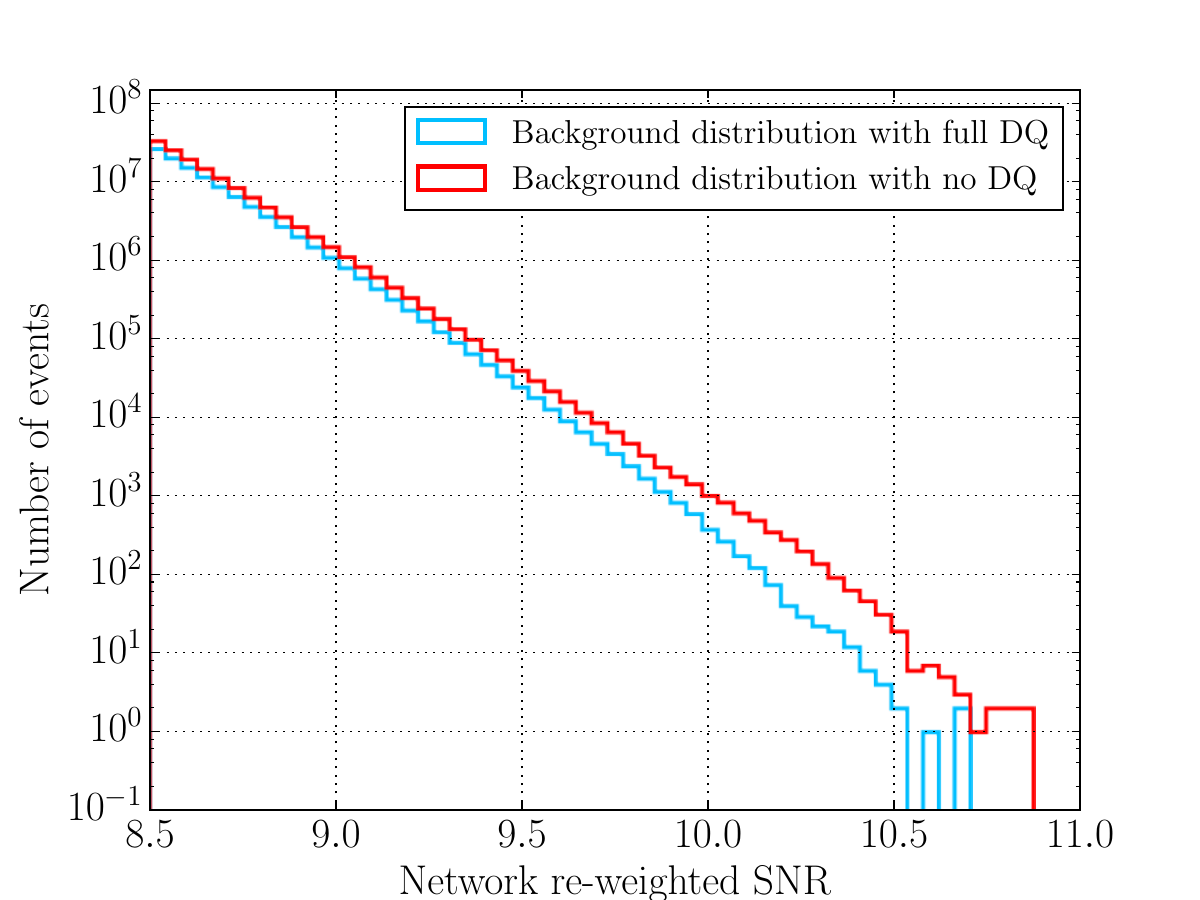}
  \label{subfig:bns-bin-gw151226-raw}
  }
  \caption{The background distribution in the BNS bin before and after applying DQ vetoes %
           for the analysis containing GW151226. %
           (\ref{subfig:bns-bin-gw151226-rate}) The cumulative rate of background triggers %
           in the BNS bin as a function of re-weighted SNR. %
           (\ref{subfig:bns-bin-gw151226-raw}) A histogram of background triggers %
           in the BNS bin. % 
           The red traces indicate the %
           distribution of background triggers without noisy data removed % 
           and the cyan traces indicate the distribution %
           of background triggers with all vetoes applied. %
          }
\label{fig:bns-bin-far-GW151226}
\end{figure}

As expected from Figure \ref{fig:GW151226-VT-RATIO}, there is a small improvement 
in the BNS background distribution when DQ vetoes are applied. 
Figure \ref{fig:bns-bin-far-GW151226} shows the background distributions in the BNS 
bin with and without noisy data removed. 
Although the $\hat{\rho}_{c}$ of the loudest background event does not 
change considerably, 
there is a noticeable gap between the two background distributions 
that is visible at $\hat{\rho}_{c} >$ 9.7 and widens to an order of 
magnitude difference in FAR at $\hat{\rho}_{c} \approx$ 10.5.

\subsection{Bulk bin}

\begin{figure}[!ht]%
\centering
  \subfloat[]{
  \includegraphics[width=0.495\textwidth]{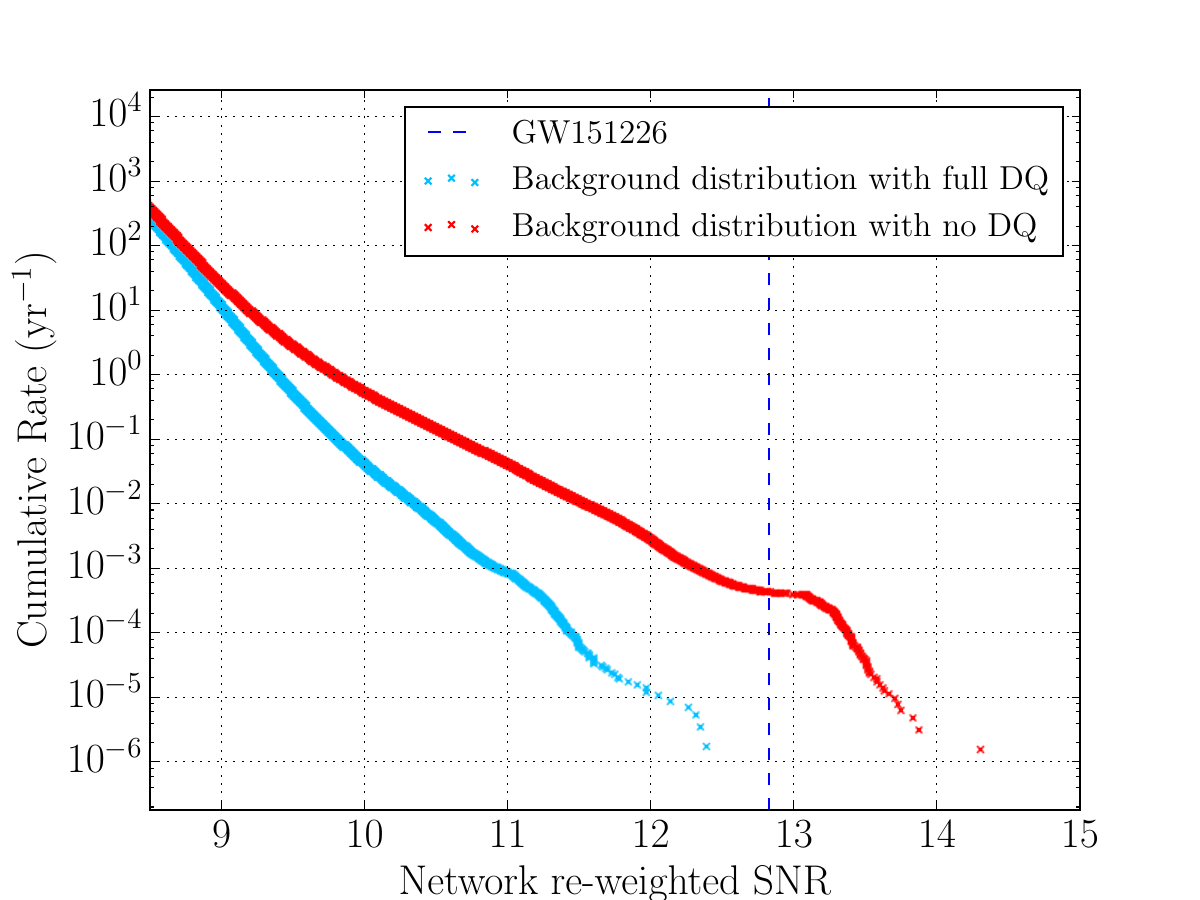}
  \label{subfig:bulk-bin-gw151226-rate}
  }
  \subfloat[]{
  \includegraphics[width=0.495\textwidth]{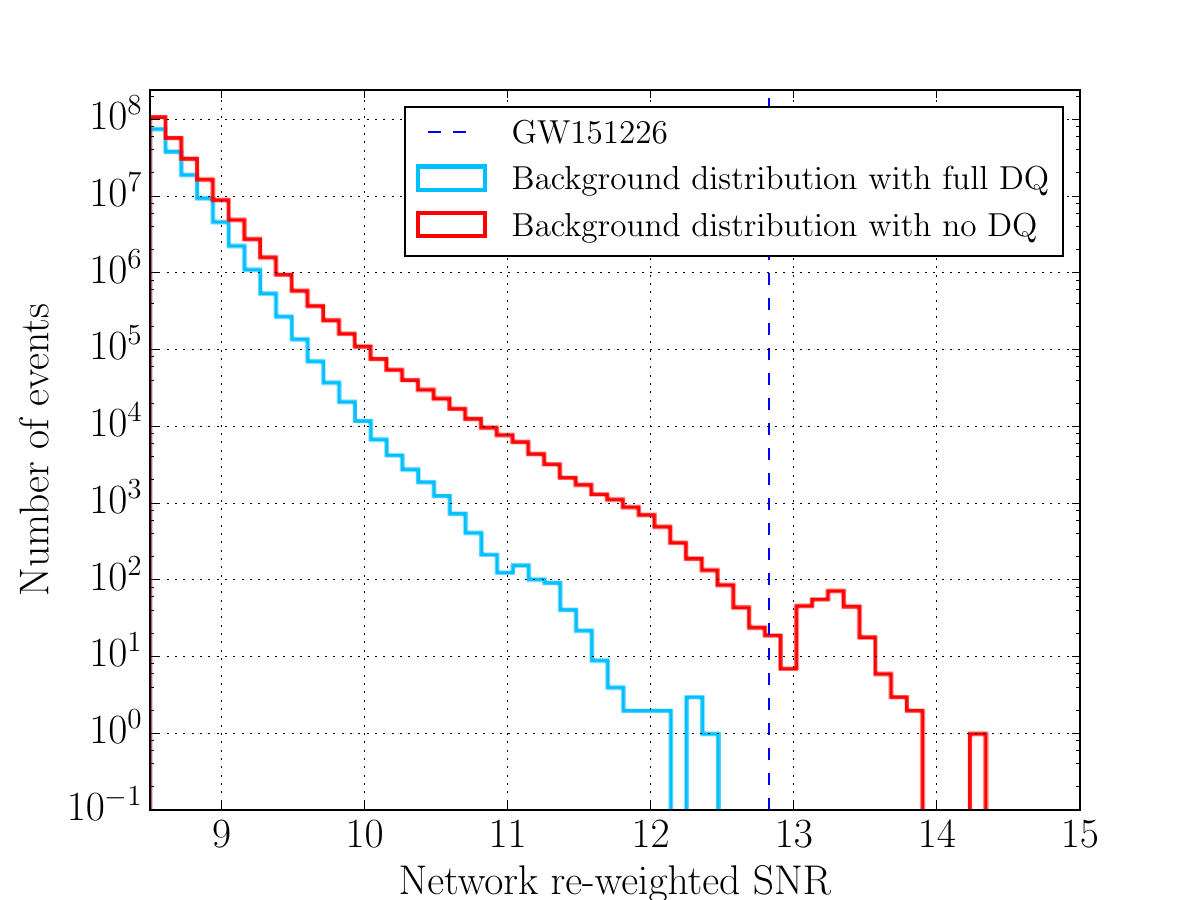}
  \label{subfig:bulk-bin-gw151226-raw}
  }
  \caption{The background distribution in the bulk bin before and after applying %
           DQ vetoes for the analysis containing GW151226. % 
           (\ref{subfig:bulk-bin-gw151226-rate}) The cumulative rate of background triggers %
           in the bulk bin as a function of re-weighted SNR. %
           (\ref{subfig:bulk-bin-gw151226-raw}) A histogram of background triggers %
           in the bulk bin. % 
           The red traces indicate the %
           distribution of background triggers with no data removed from the analysis. %
           The cyan traces indicate the distribution %
           of background triggers with all DQ vetoes applied. %
           The dotted line indicates GW151226, which was recovered with $\hat{\rho}_{c} =$ 12.7. %
           If DQ vetoes are not applied, GW151226 is no%
           longer louder than the entire background distribution. %
          }
\label{fig:bulk-bin-far-GW151226}
\end{figure}

The background distribution in the bulk bin changes significantly when DQ vetoes 
are applied, which is shown in Figure \ref{fig:bulk-bin-far-GW151226}. There is a 
visible difference between the two distributions beginning at $\hat{\rho}_{c} =$ 9. 
The difference in cumulative rate reaches an order of magnitude at 
$\hat{\rho}_{c} \sim$ 10 and continues to grow for larger values of $\hat{\rho}_{c}$. 
The reduction of the background distribution through the application of DQ vetoes is 
particularly impactful for GW151226, which is discussed in Section \ref{sec:GW151226}.

\subsubsection{GW151226}\label{sec:GW151226}

The second binary black hole system discovered in the first observing run, 
GW151226 \cite{GW151226}, is indicated by the vertical dotted line at $\hat{\rho}_{c} =$ 12.7 in 
Figure \ref{fig:bulk-bin-far-GW151226}. When noisy data are removed from the analysis, 
the background distribution in the bulk bin is reduced and GW151226 is the loudest 
event in the analysis. The false alarm rate of GW151226 decreases by over two 
orders of magnitude, resulting in a clear detection. The false alarm rates before 
and after DQ vetoes are applied are listed in Table \ref{table:GW151226-far}.

\begin{table}[!ht]%
  \begin{center}
    \begin{tabular}{lc}
      \hline
      Analysis configuration & False alarm rate ($\mathrm{yr}^{-1}$) \\ \hline
      All vetoes applied & $<~5.39\times10^{-6}$ \\ 
      No vetoes applied & $1.30\times10^{-3}$ \\
      \hline 
    \end{tabular}
  \end{center}
  \caption{Table of bulk bin false alarm rates of GW151226. %
           The false alarm rate of GW151226 increases %
           from less than 1 in 186000 years to 1 in 770 years if data with excess noise %
           is not removed from the analysis.}
  \label{table:GW151226-far}
\end{table}

\subsection{Edge bin}

\begin{figure}[!ht]%
\centering
  \subfloat[]{
  \includegraphics[width=0.495\textwidth]{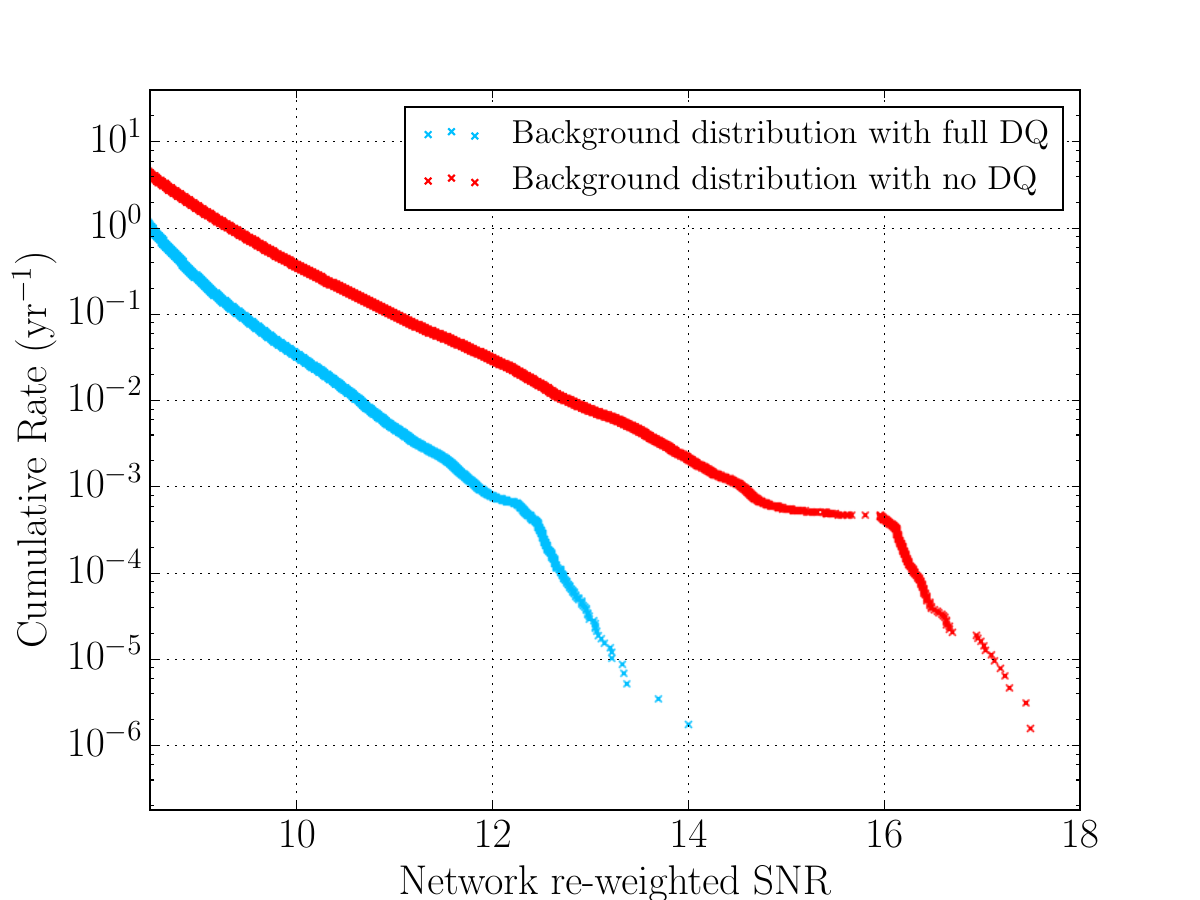}
  \label{subfig:edge-bin-gw151226-rate}
  }
  \subfloat[]{
  \includegraphics[width=0.495\textwidth]{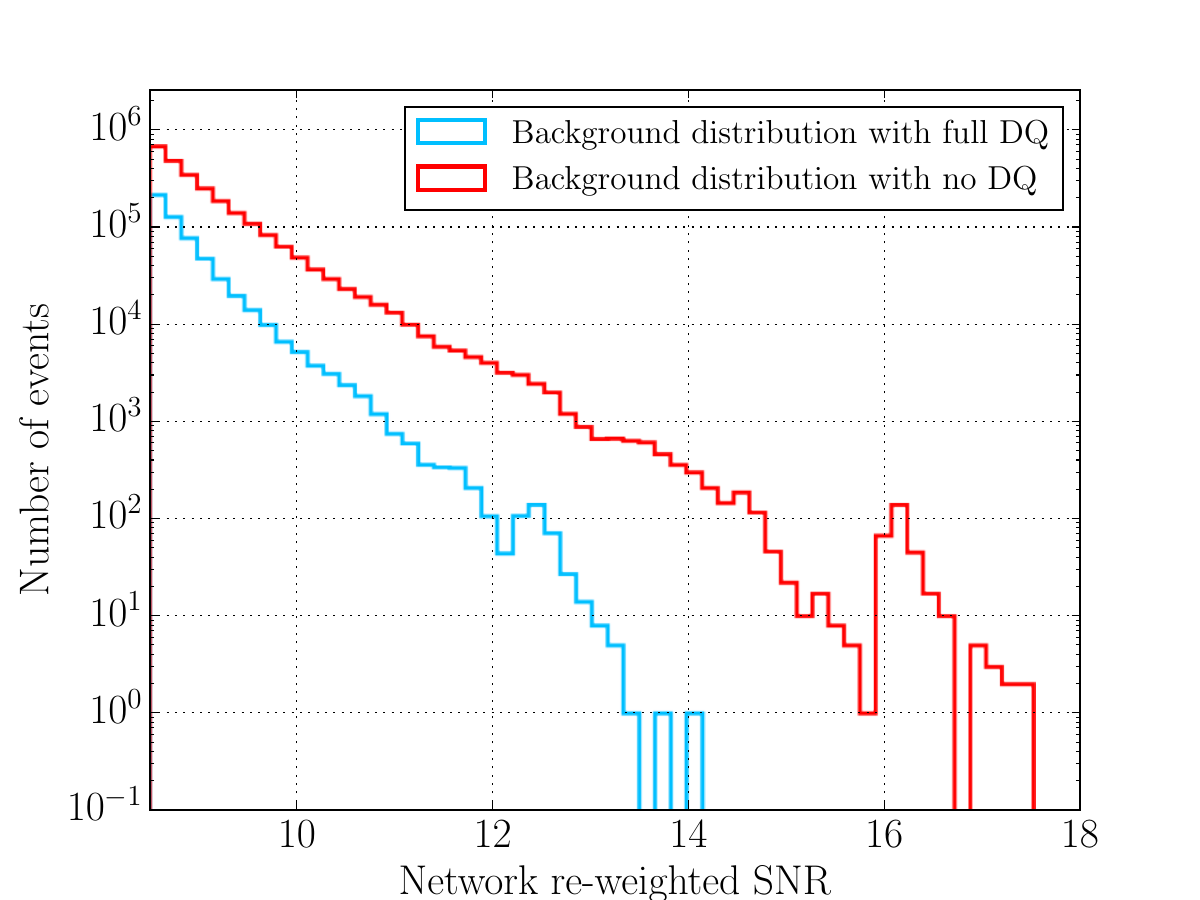}
  \label{subfig:edge-bin-gw151226-raw}
  }
  \caption{The background distribution in the edge bin before and after applying DQ vetoes %
           for the analysis containing GW151226. %
           (\ref{subfig:edge-bin-gw151226-rate}) The cumulative rate of background triggers %
           in the edge bin as a function of re-weighted SNR. %
           (\ref{subfig:edge-bin-gw151226-raw}) A histogram of background triggers %
           in the bulk bin. % 
           The red traces indicate the distribution of background triggers without %
           removing noisy data and the cyan traces indicate the distribution %
           of background triggers with all vetoes applied. %
          }
\label{fig:edge-bin-far-GW151226}
\end{figure}

Figure \ref{fig:edge-bin-far-GW151226} shows the background 
distribution in the edge bin before and after DQ vetoes have been applied. 
The background distribution of the edge bin 
in the analysis containing GW151226 is significantly reduced for all values of $\hat{\rho}_{c}$ 
when DQ vetoes are applied.  

\section{Limiting noise sources}\label{sec:limiting}

The sensitivity of the search is limited by instrumental features that result in high 
$\hat{\rho}_{c}$ triggers and tails in the background distributions.
This section uses two particular instrumental noise sources from the analysis 
containing GW150914 as case studies. These are examples of noise sources that 
are not able to be vetoed using existing algorithms as they are not captured by 
any existing witness sensors in the detectors.

\subsection{Blip transients}

\begin{figure}[!ht]%
\centering
  \includegraphics[width=0.75\textwidth]{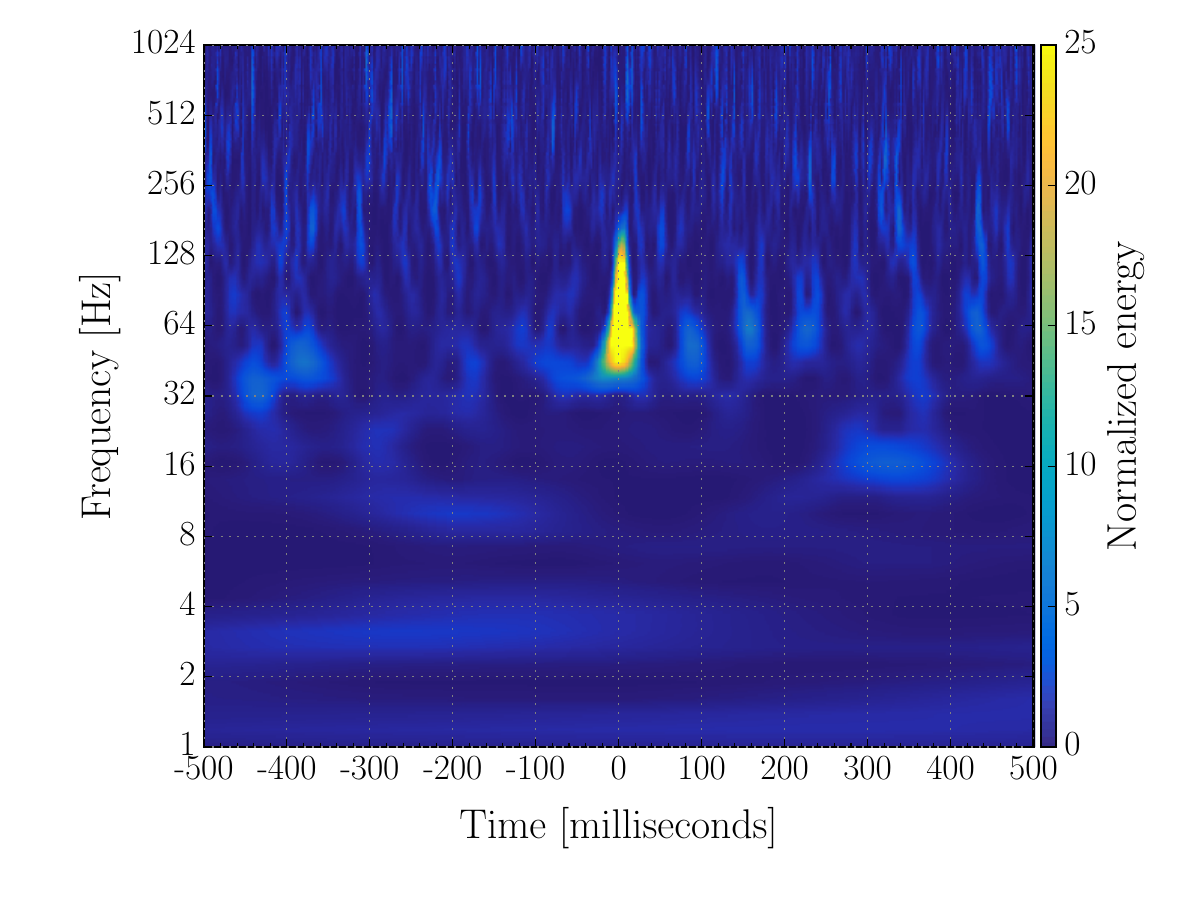}
  \caption{A time-frequency representation \cite{ref:omegagrams} of the Livingston strain channel %
           at the time of a blip transient. %
           This visualization of a blip transient demonstrates their typical %
           features: band-limited, short duration, and little visible frequency structure.}
\label{fig:Blip-omega}
\end{figure}

Blip transients \cite{GW150914-DETCHAR} are band-limited noise transients that occur 
in both the Hanford and Livingston detectors. They do not occur in coincidence between 
the two LIGO detectors and are not candidate gravitational wave signals. 
Due to their short duration in LIGO's 
sensitive frequency band, the $\chi^{2}$ test is not as effective at down-weighting 
these noise transients and they have been a common source of high re-weighted SNR background 
triggers. Figure \ref{fig:Blip-omega} shows a time-frequency representation of a 
blip transient. 

\begin{figure}[!ht]%
\centering
  \includegraphics[width=\textwidth]{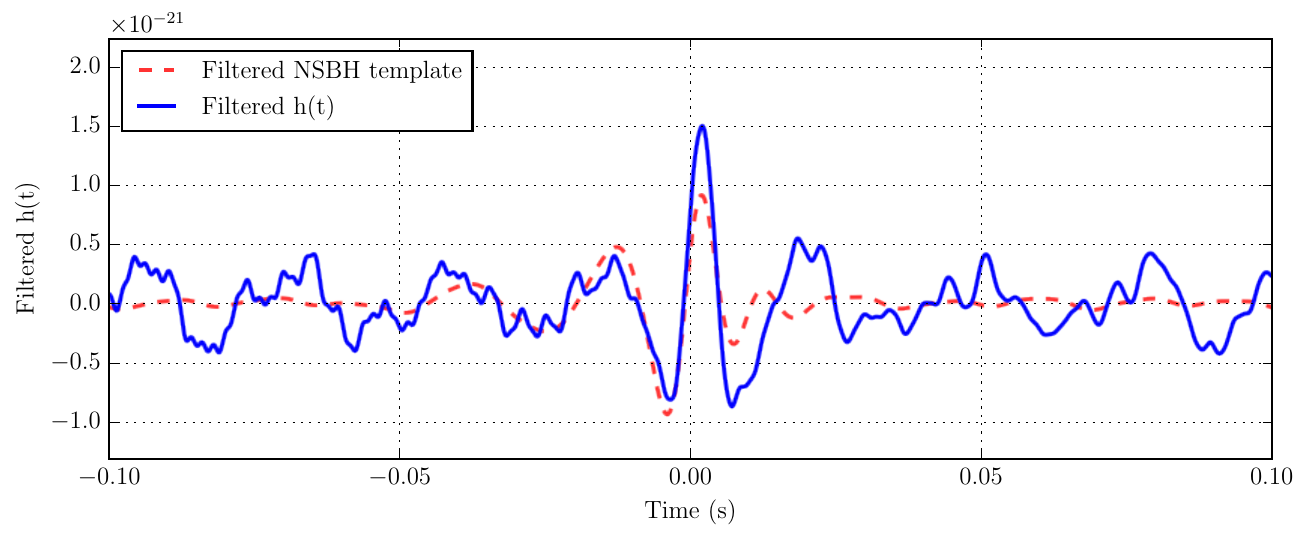}
  \caption{A filtered time-domain representation of the Livingston strain channel $h(t)$ at %
           the time of a blip transient. 
           The dotted line is a filtered CBC waveform that reported %
           a high re-weighted SNR value at the time of the blip transient. Both sets of data %
           have been zero-phase bandpass filtered to isolate the frequency range that aLIGO %
           is sensitive to. The short duration and high overlap of %
           these two curves causes the $\chi^{2}$ to be ineffective at down-weighting these transients.}
\label{fig:Blip-cbc-waveform}
\end{figure}

Figure \ref{fig:Blip-cbc-waveform} shows a 
time-domain representation of a blip transient in the Livingston strain channel. 
The dotted line on top of the strain data represents a template waveform that reported a 
high re-weighted SNR at the time of this blip transient. Both curves have been filtered 
to isolate the parts of the signal that are in LIGO's sensitive frequency band. The 
template waveform is a high mass system $M_{\mathrm{total}} = 98.34~\mathrm{M_{\odot}}$) 
with an anti-aligned effective spin of $-0.97$. This waveform is among the shortest 
duration templates used in the analysis, spending less than 0.1 seconds in LIGO's 
sensitive frequency band.

\begin{figure}[!ht]%
\centering
  \subfloat[]{
  \includegraphics[width=0.5\textwidth]{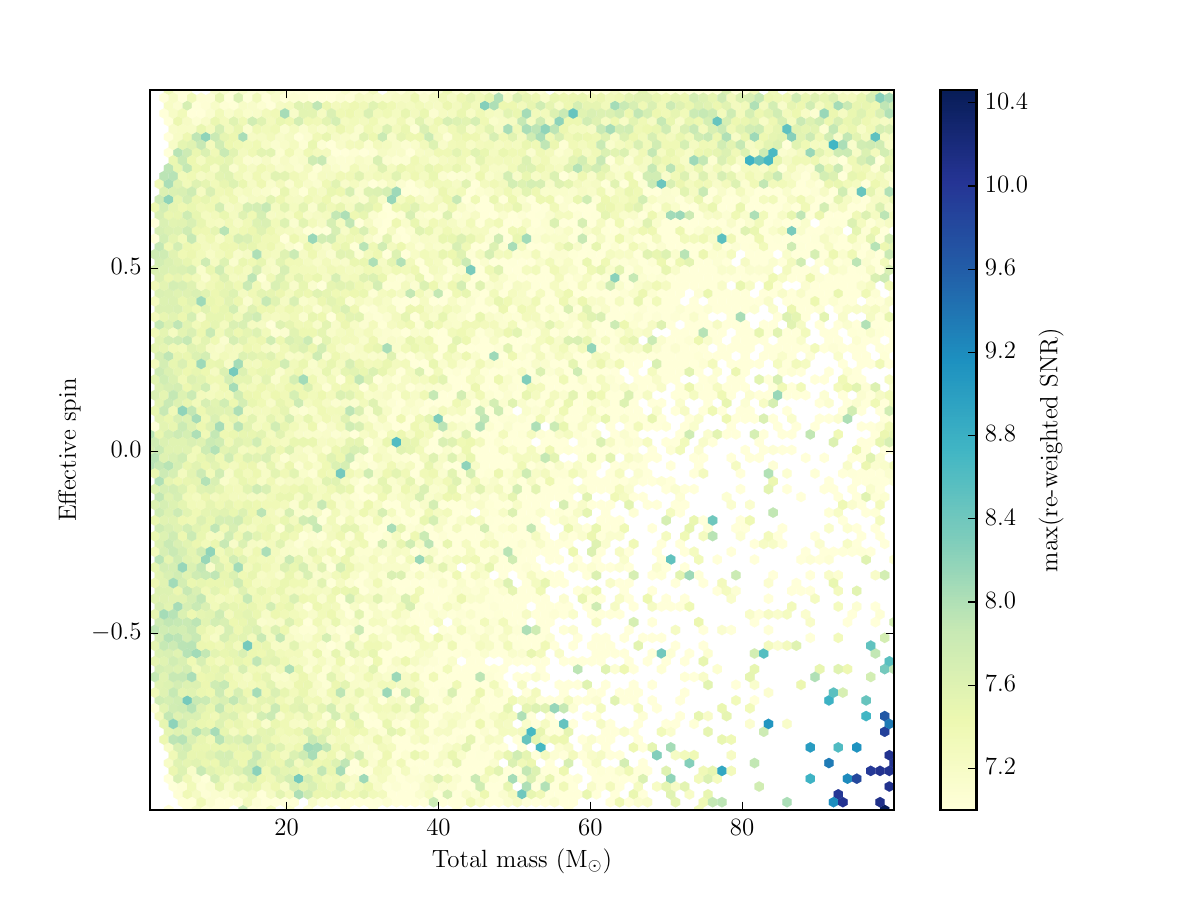}
  \label{fig:GW150914_mtotal_seff_modot}
  }
  \subfloat[]{
  \includegraphics[width=0.5\textwidth]{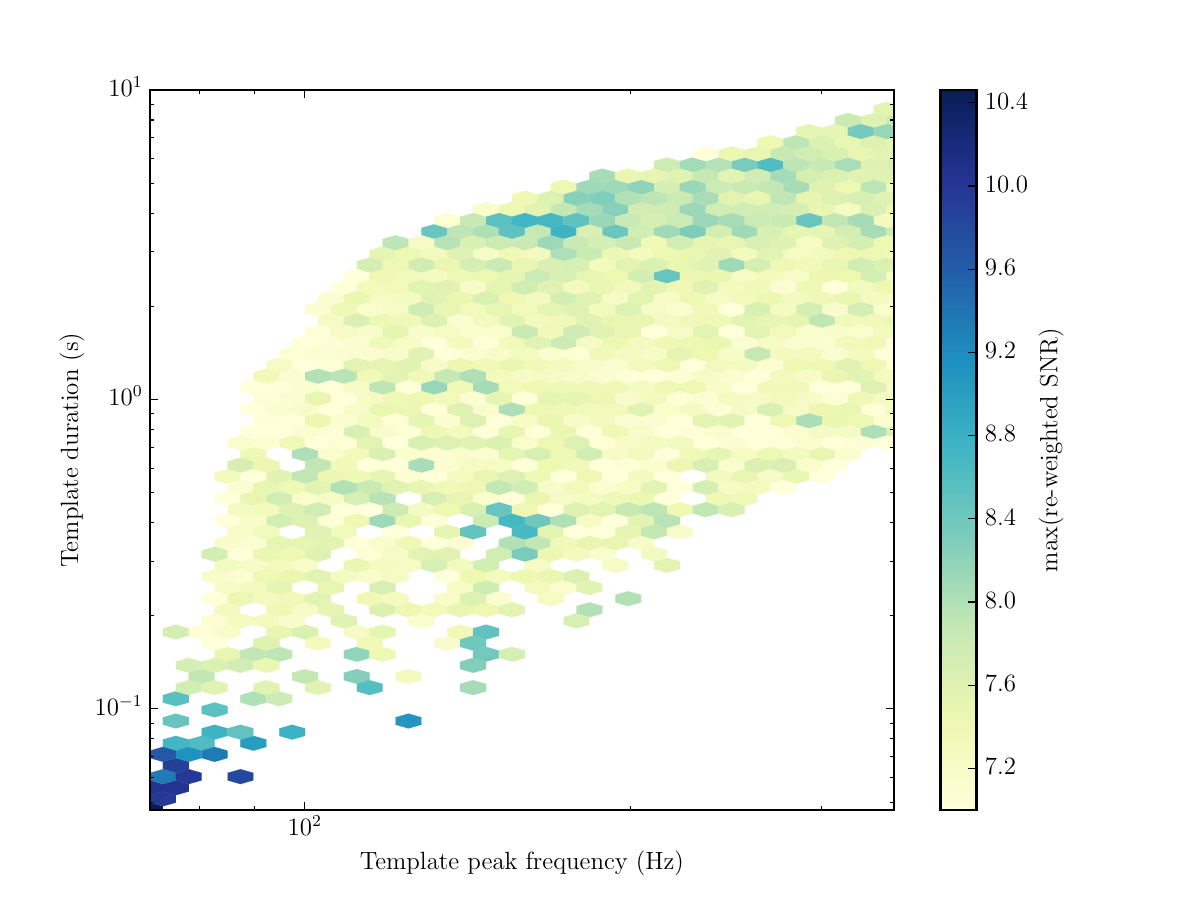}
  \label{fig:fpeak-template-duration-hex}
  }
  \caption{Single detector triggers from the Livingston detector during the analysis containing %
           GW150914. %
           (\ref{fig:GW150914_mtotal_seff_modot}) Triggers binned by total mass and effective spin. %
           The highest re-weighted SNR triggers are constrained to the bottom corner of the plot, bounded by %
           $M_{\mathrm{total}} > 80$ and $\chi_{\mathrm{eff}} < -0.5$. %
           (\ref{fig:fpeak-template-duration-hex}) Triggers binned by the peak frequency and duration % 
           of the template waveform. The loudest triggers in re-weighted SNR are %
           constrained to the area of the parameter space with template durations $<$ 0.1 seconds. %
           The small cluster of loud triggers with a template duration of roughly 4 seconds corresponds to %
           the 60-200 Hz noise discussed in Section \ref{sec:60-200-hz-noise}.%
          }
\label{fig:trigger-hexbins}
\end{figure}

The region of the astrophysical parameter space where the $\chi^{2}$ test is ineffective at 
down-weighting blip transients is small. This is demonstrated in Figure \ref{fig:trigger-hexbins}, 
which shows triggers from the Livingston detector during the analysis containing GW150914. Each 
point represents the highest single detector re-weighted SNR measured in that region of the 
parameter space. Figure \ref{fig:GW150914_mtotal_seff_modot} shows this trigger set binned by 
total mass and effective spin. Templates that overlap with blip transients and produce high 
re-weighted SNR triggers, such as the template plotted in Figure \ref{fig:Blip-cbc-waveform}, 
are constrained to the region where $M_{\mathrm{total}} > 80 \mathrm{M_{\odot}}$ and 
$\chi_{\mathrm{eff}} < -0.5$. This region contains only 65 waveform templates out of 249077 
total waveforms in the template bank; the majority of the template bank is capable of rejecting 
blip glitches using the $\chi^{2}$ test. Blip glitches that cause loud background events in this 
region of the template bank populate the tail of the background distribution in the edge bin
and limit search sensitivity in that bin.

The region of the template bank where the $\chi^{2}$ test is ineffective can also be 
visualized in terms of the duration of the template waveform. Figure 
\ref{fig:fpeak-template-duration-hex} shows the same trigger set binned by the peak frequency 
and duration of the template waveform. In this view of the parameter space, the templates with 
the highest re-weighted SNR are contrained to the region where the duration of the template  
in LIGO's sensitive frequency band is less than 0.1 seconds. In this region, the duration 
of the template is on the same time scale as instrumental transients such as blip transients.

Mitigation of blip transients is a high priority but is difficult since they do not couple into instrumental 
witness sensors 
and are not high enough in amplitude to be removed by the gating process. Since an instrumental cause 
has not yet been identified, a modified ranking 
statistic \cite{blip-rejector} that is capable of better discriminating blip transients from 
gravitational wave signals has been developed and implemented for the second observing run. 

\subsection{60-200 Hz noise}\label{sec:60-200-hz-noise}

\begin{figure}[!ht]%
\centering
  \subfloat[]{
      \includegraphics[width=\textwidth]{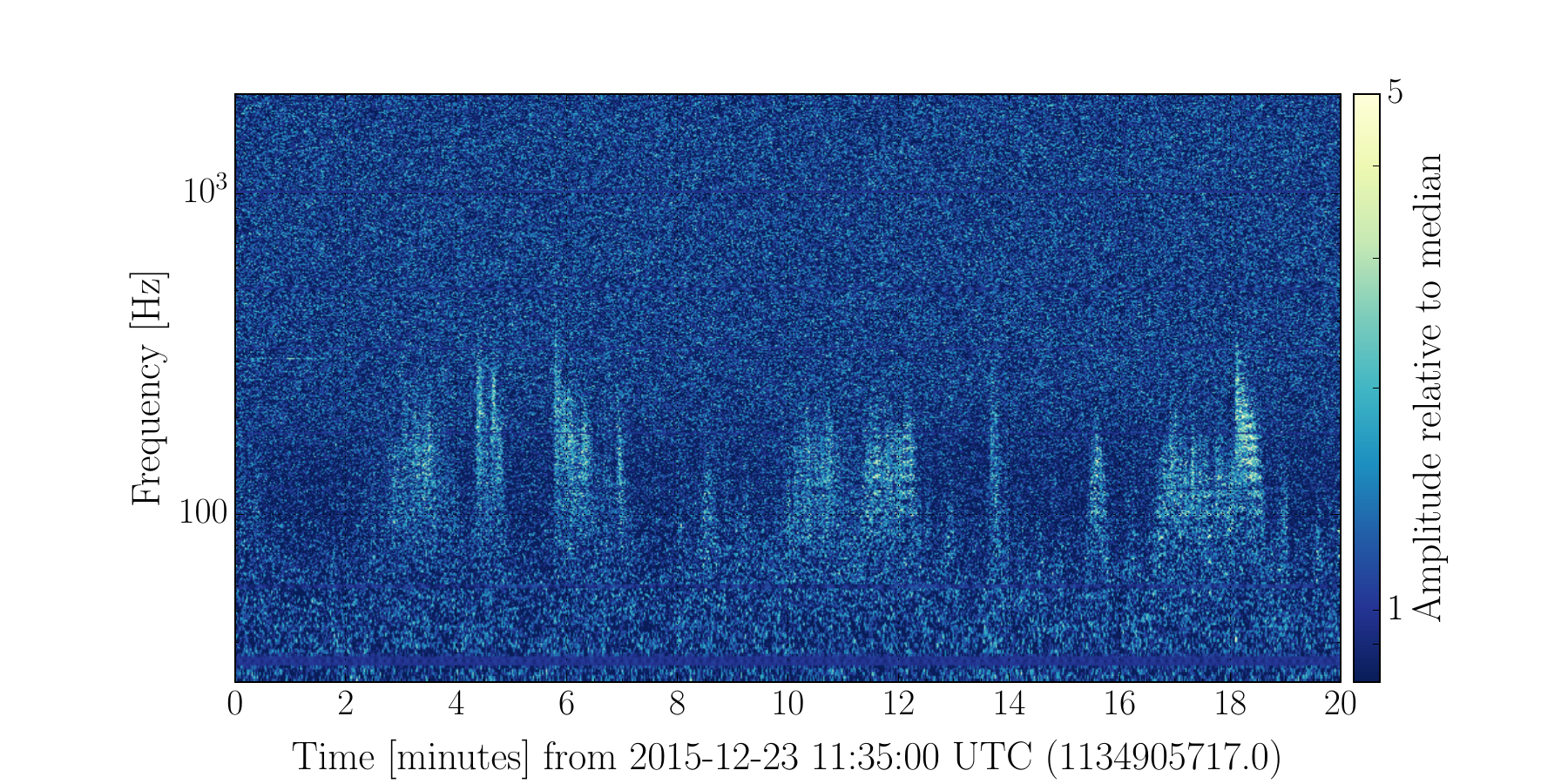}
      \label{fig:60-200-Hz-noise}
  }

  \subfloat[]{
      \includegraphics[width=\textwidth]{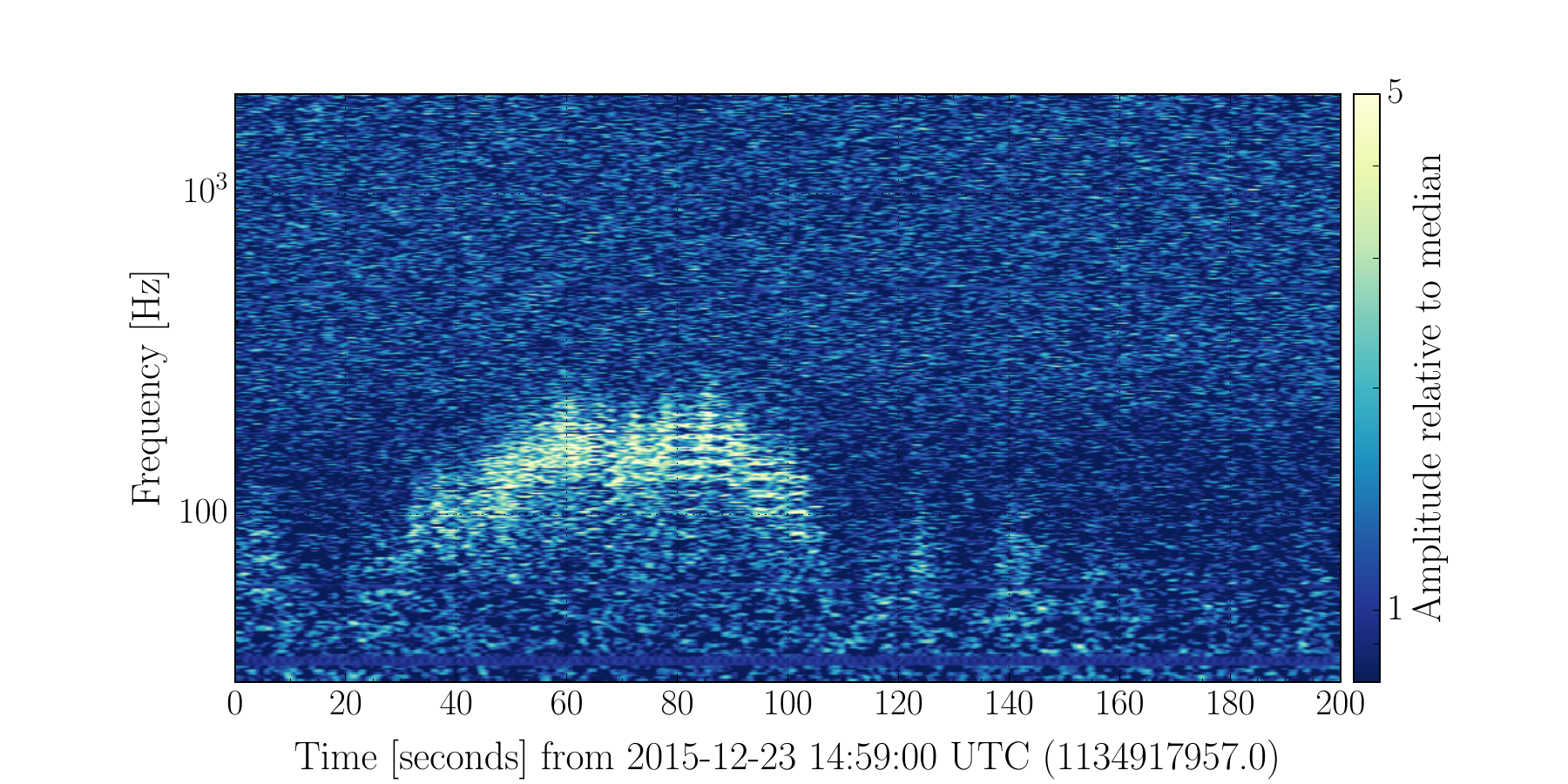}
      \label{fig:zoom-60-200-Hz-noise}
  }

  \caption{Time-frequency spectrograms of the 60-200 Hz noise. %
           (\ref{fig:60-200-Hz-noise}) A 20 minute time scale shows the 60-200 Hz noise appearing %
           for several minutes at a time. This time scale and frequency range is damaging %
           to CBC searches and has often been found responsible for loud background events. %
           (\ref{fig:zoom-60-200-Hz-noise}) A 200 second time scale reveals the arc-like %
           shape of the noise in the time-frequency plane. % 
           This period of noise caused a loud trigger in the PyCBC background. % 
  }  
  \label{fig:60-200-Hz-both}
\end{figure}

A second problematic noise source that was present at Livingston during the first observing run 
was the ``60-200 Hz'' noise. Figure \ref{fig:60-200-Hz-both} shows a time-frequency visualization 
of this noise on both 200 second and 20 minute time scales. This noise source appears for several 
minutes at a time as arc-like patterns in the time-frequency plane. This noise source causes 
loud background triggers when analyzing the Livingston data, including the band of triggers 
with a template duration of $\sim$4 seconds in Figure \ref{fig:fpeak-template-duration-hex}.
Although the arc-like patterns in Figure \ref{fig:zoom-60-200-Hz-noise} are similar to those 
caused by scattered light \cite{virgo-scattering}, the exact cause of this noise is not yet 
fully understood.

\section{Conclusions}

Data quality vetoes improved the sensitivity of the PyCBC search in Advanced LIGO's first 
observing run. Although the sensitivity of the search to BNS signals was not dramatically 
affected, VT improved significantly for higher mass sources when DQ vetoes are applied. 

The gravitational wave signal GW150914 was strong enough that it was louder than all background 
events regardless of what data were removed from the search. As such, DQ vetoes did not 
improve its significance. 
The false alarm rate of LVT151012, which occurred during the same analysis period, was 
improved from 0.69 $\mathrm{yr}^{-1}$ to 0.33 $\mathrm{yr}^{-1}$ when 
vetoes were applied. 
The false alarm rate of the second gravitational wave signal discovered in O1, GW151226, was 
decreased by over two orders of magnitude when DQ vetoes were applied, which 
resulted in a clear detection. 
The production and application of DQ vetoes was critical for increasing overall sensitivity 
in Advanced LIGO's first observing run and similar methods were employed during the 
second observing run.

\section{Acknowledgements}
\input{LVCacknowledgements.tex}

\section{References}
\bibliographystyle{unsrt}
\bibliography{updated_refs,cbc-group,GW150914_refs}

\end{document}

%% file: LSC_Feb2016_Virgo_Feb2016-cqg.tex
%% LSC authorlist in CQG format
%\documentclass[12pt]{iopart}
%\begin{document}
%\title{LSC February 2016 and Virgo Febuary 2016 author list---LIGO-M1600028\\
%10/10/2017. CQG style}
\author{%
B~P~Abbott$^{1}$,  %benjamin.abbott
R~Abbott$^{1}$,  %rich.abbott
T~D~Abbott$^{2}$,  %thomas.abbott
M~R~Abernathy$^{3}$,  %matthew.abernathy
F~Acernese$^{4,5}$, %fausto.acernese
K~Ackley$^{6}$,  %kendall.ackley
C~Adams$^{7}$,  %carl.adams
T~Adams$^{8}$, %thomas.adams
P~Addesso$^{9}$,  %paolo.addesso
R~X~Adhikari$^{1}$
% ...
\\
{(LIGO Scientific Collaboration and Virgo Collaboration)}%
}%
\medskip
\address {${}^{*}$Deceased, March 2016. ${}^{\dag}$Deceased, May 2015. ${}^{\ddag}$Deceased, March 2015. }% 
\medskip
\address {$^{1}$LIGO, California Institute of Technology, Pasadena, CA 91125, USA }
\address {$^{2}$Louisiana State University, Baton Rouge, LA 70803, USA }
\address {$^{3}$American University, Washington, D.C. 20016, USA }
\address {$^{4}$Universit\`a di Salerno, Fisciano, I-84084 Salerno, Italy }
\address {$^{5}$INFN, Sezione di Napoli, Complesso Universitario di Monte S.Angelo, I-80126 Napoli, Italy }
\address {$^{6}$University of Florida, Gainesville, FL 32611, USA }
\address {$^{7}$LIGO Livingston Observatory, Livingston, LA 70754, USA }
\address {$^{8}$Laboratoire d'Annecy-le-Vieux de Physique des Particules (LAPP), Universit\'e Savoie Mont Blanc, CNRS/IN2P3, F-74941 Annecy-le-Vieux, France }
\address {$^{9}$University of Sannio at Benevento, I-82100 Benevento, Italy and INFN, Sezione di Napoli, I-80100 Napoli, Italy }
\address {$^{10}$Albert-Einstein-Institut, Max-Planck-Institut f\"ur Gravi\-ta\-tions\-physik, D-30167 Hannover, Germany }
\address {$^{11}$Nikhef, Science Park, 1098 XG Amsterdam, The Netherlands }
\address {$^{12}$LIGO, Massachusetts Institute of Technology, Cambridge, MA 02139, USA }
\address {$^{13}$Instituto Nacional de Pesquisas Espaciais, 12227-010 S\~{a}o Jos\'{e} dos Campos, S\~{a}o Paulo, Brazil }
\address {$^{14}$INFN, Gran Sasso Science Institute, I-67100 L'Aquila, Italy }
\address {$^{15}$INFN, Sezione di Roma Tor Vergata, I-00133 Roma, Italy }
\address {$^{16}$Inter-University Centre for Astronomy and Astrophysics, Pune 411007, India }
\address {$^{17}$University of Wisconsin-Milwaukee, Milwaukee, WI 53201, USA }
\address {$^{18}$Leibniz Universit\"at Hannover, D-30167 Hannover, Germany }
\address {$^{19}$Universit\`a di Pisa, I-56127 Pisa, Italy }
\address {$^{20}$INFN, Sezione di Pisa, I-56127 Pisa, Italy }
\address {$^{21}$Australian National University, Canberra, Australian Capital Territory 0200, Australia }
\address {$^{22}$The University of Mississippi, University, MS 38677, USA }
\address {$^{23}$California State University Fullerton, Fullerton, CA 92831, USA }
\address {$^{24}$LAL, Univ. Paris-Sud, CNRS/IN2P3, Universit\'e Paris-Saclay, Orsay, France }
\address {$^{25}$Chennai Mathematical Institute, Chennai 603103, India }
\address {$^{26}$Universit\`a di Roma Tor Vergata, I-00133 Roma, Italy }
\address {$^{27}$University of Southampton, Southampton SO17 1BJ, United Kingdom }
\address {$^{28}$Universit\"at Hamburg, D-22761 Hamburg, Germany }
\address {$^{29}$INFN, Sezione di Roma, I-00185 Roma, Italy }
\address {$^{30}$Albert-Einstein-Institut, Max-Planck-Institut f\"ur Gravitations\-physik, D-14476 Potsdam-Golm, Germany }
\address {$^{31}$APC, AstroParticule et Cosmologie, Universit\'e Paris Diderot, CNRS/IN2P3, CEA/Irfu, Observatoire de Paris, Sorbonne Paris Cit\'e, F-75205 Paris Cedex 13, France }
\address {$^{32}$Montana State University, Bozeman, MT 59717, USA }
\address {$^{33}$Universit\`a di Perugia, I-06123 Perugia, Italy }
\address {$^{34}$INFN, Sezione di Perugia, I-06123 Perugia, Italy }
\address {$^{35}$European Gravitational Observatory (EGO), I-56021 Cascina, Pisa, Italy }
\address {$^{36}$Syracuse University, Syracuse, NY 13244, USA }
\address {$^{37}$SUPA, University of Glasgow, Glasgow G12 8QQ, United Kingdom }
\address {$^{38}$LIGO Hanford Observatory, Richland, WA 99352, USA }
\address {$^{39}$Wigner RCP, RMKI, H-1121 Budapest, Konkoly Thege Mikl\'os \'ut 29-33, Hungary }
\address {$^{40}$Columbia University, New York, NY 10027, USA }
\address {$^{41}$Stanford University, Stanford, CA 94305, USA }
\address {$^{42}$Universit\`a di Padova, Dipartimento di Fisica e Astronomia, I-35131 Padova, Italy }
\address {$^{43}$INFN, Sezione di Padova, I-35131 Padova, Italy }
\address {$^{44}$CAMK-PAN, 00-716 Warsaw, Poland }
\address {$^{45}$University of Birmingham, Birmingham B15 2TT, United Kingdom }
\address {$^{46}$Universit\`a degli Studi di Genova, I-16146 Genova, Italy }
\address {$^{47}$INFN, Sezione di Genova, I-16146 Genova, Italy }
\address {$^{48}$RRCAT, Indore MP 452013, India }
\address {$^{49}$Faculty of Physics, Lomonosov Moscow State University, Moscow 119991, Russia }
\address {$^{50}$SUPA, University of the West of Scotland, Paisley PA1 2BE, United Kingdom }
\address {$^{51}$University of Western Australia, Crawley, Western Australia 6009, Australia }
\address {$^{52}$Department of Astrophysics/IMAPP, Radboud University Nijmegen, P.O. Box 9010, 6500 GL Nijmegen, The Netherlands }
\address {$^{53}$Artemis, Universit\'e C\^ote d'Azur, CNRS, Observatoire C\^ote d'Azur, CS 34229, Nice cedex 4, France }
\address {$^{54}$Institut de Physique de Rennes, CNRS, Universit\'e de Rennes 1, F-35042 Rennes, France }
\address {$^{55}$Washington State University, Pullman, WA 99164, USA }
\address {$^{56}$Universit\`a degli Studi di Urbino ``Carlo Bo,'' I-61029 Urbino, Italy }
\address {$^{57}$INFN, Sezione di Firenze, I-50019 Sesto Fiorentino, Firenze, Italy }
\address {$^{58}$University of Oregon, Eugene, OR 97403, USA }
\address {$^{59}$Laboratoire Kastler Brossel, UPMC-Sorbonne Universit\'es, CNRS, ENS-PSL Research University, Coll\`ege de France, F-75005 Paris, France }
\address {$^{60}$Carleton College, Northfield, MN 55057, USA }
\address {$^{61}$Astronomical Observatory Warsaw University, 00-478 Warsaw, Poland }
\address {$^{62}$VU University Amsterdam, 1081 HV Amsterdam, The Netherlands }
\address {$^{63}$University of Maryland, College Park, MD 20742, USA }
\address {$^{64}$Center for Relativistic Astrophysics and School of Physics, Georgia Institute of Technology, Atlanta, GA 30332, USA }
\address {$^{65}$Laboratoire des Mat\'eriaux Avanc\'es (LMA), CNRS/IN2P3, F-69622 Villeurbanne, France }
\address {$^{66}$Universit\'e Claude Bernard Lyon 1, F-69622 Villeurbanne, France }
\address {$^{67}$Universit\`a di Napoli ``Federico II,'' Complesso Universitario di Monte S.Angelo, I-80126 Napoli, Italy }
\address {$^{68}$NASA/Goddard Space Flight Center, Greenbelt, MD 20771, USA }
\address {$^{69}$RESCEU, University of Tokyo, Tokyo, 113-0033, Japan. }
\address {$^{70}$Tsinghua University, Beijing 100084, China }
\address {$^{71}$Texas Tech University, Lubbock, TX 79409, USA }
\address {$^{72}$The Pennsylvania State University, University Park, PA 16802, USA }
\address {$^{73}$National Tsing Hua University, Hsinchu City, 30013 Taiwan, Republic of China }
\address {$^{74}$Charles Sturt University, Wagga Wagga, New South Wales 2678, Australia }
\address {$^{75}$West Virginia University, Morgantown, WV 26506, USA }
\address {$^{76}$University of Chicago, Chicago, IL 60637, USA }
\address {$^{77}$Caltech CaRT, Pasadena, CA 91125, USA }
\address {$^{78}$Korea Institute of Science and Technology Information, Daejeon 305-806, Korea }
\address {$^{79}$Universit\`a di Roma ``La Sapienza,'' I-00185 Roma, Italy }
\address {$^{80}$University of Brussels, Brussels 1050, Belgium }
\address {$^{81}$Sonoma State University, Rohnert Park, CA 94928, USA }
\address {$^{82}$Center for Interdisciplinary Exploration \& Research in Astrophysics (CIERA), Northwestern University, Evanston, IL 60208, USA }
\address {$^{83}$University of Minnesota, Minneapolis, MN 55455, USA }
\address {$^{84}$The University of Melbourne, Parkville, Victoria 3010, Australia }
\address {$^{85}$Institute for Plasma Research, Bhat, Gandhinagar 382428, India }
\address {$^{86}$The University of Sheffield, Sheffield S10 2TN, United Kingdom }
\address {$^{87}$The University of Texas Rio Grande Valley, Brownsville, TX 78520, USA }
\address {$^{88}$Universit\`a di Trento, Dipartimento di Fisica, I-38123 Povo, Trento, Italy }
\address {$^{89}$INFN, Trento Institute for Fundamental Physics and Applications, I-38123 Povo, Trento, Italy }
\address {$^{90}$Cardiff University, Cardiff CF24 3AA, United Kingdom }
\address {$^{91}$Montclair State University, Montclair, NJ 07043, USA }
\address {$^{92}$MTA E\"otv\"os University, ``Lendulet'' Astrophysics Research Group, Budapest 1117, Hungary }
\address {$^{93}$National Astronomical Observatory of Japan, 2-21-1 Osawa, Mitaka, Tokyo 181-8588, Japan }
\address {$^{94}$School of Mathematics, University of Edinburgh, Edinburgh EH9 3FD, United Kingdom }
\address {$^{95}$Indian Institute of Technology, Gandhinagar Ahmedabad Gujarat 382424, India }
\address {$^{96}$University of Szeged, D\'om t\'er 9, Szeged 6720, Hungary }
\address {$^{97}$International Centre for Theoretical Sciences, Tata Institute of Fundamental Research, Bangalore 560012, India }
\address {$^{98}$Embry-Riddle Aeronautical University, Prescott, AZ 86301, USA }
\address {$^{99}$Tata Institute of Fundamental Research, Mumbai 400005, India }
\address {$^{100}$INAF, Osservatorio Astronomico di Capodimonte, I-80131, Napoli, Italy }
\address {$^{101}$University of Michigan, Ann Arbor, MI 48109, USA }
\address {$^{102}$Rochester Institute of Technology, Rochester, NY 14623, USA }
\address {$^{103}$NCSA, University of Illinois at Urbana-Champaign, Urbana, Illinois 61801, USA }
\address {$^{104}$Universitat de les Illes Balears, IAC3---IEEC, E-07122 Palma de Mallorca, Spain }
\address {$^{105}$University of Bia{\l }ystok, 15-424 Bia{\l }ystok, Poland }
\address {$^{106}$SUPA, University of Strathclyde, Glasgow G1 1XQ, United Kingdom }
\address {$^{107}$IISER-TVM, CET Campus, Trivandrum Kerala 695016, India }
\address {$^{108}$Canadian Institute for Theoretical Astrophysics, University of Toronto, Toronto, Ontario M5S 3H8, Canada }
\address {$^{109}$Institute of Applied Physics, Nizhny Novgorod, 603950, Russia }
\address {$^{110}$Pusan National University, Busan 609-735, Korea }
\address {$^{111}$Hanyang University, Seoul 133-791, Korea }
\address {$^{112}$University of Adelaide, Adelaide, South Australia 5005, Australia }
\address {$^{113}$NCBJ, 05-400 \'Swierk-Otwock, Poland }
\address {$^{114}$IM-PAN, 00-956 Warsaw, Poland }
\address {$^{115}$Monash University, Victoria 3800, Australia }
\address {$^{116}$Seoul National University, Seoul 151-742, Korea }
\address {$^{117}$The Chinese University of Hong Kong, Shatin, NT, Hong Kong }
\address {$^{118}$University of Alabama in Huntsville, Huntsville, AL 35899, USA }
\address {$^{119}$University of Massachusetts-Amherst, Amherst, MA 01003, USA }
\address {$^{120}$ESPCI, CNRS, F-75005 Paris, France }
\address {$^{121}$Universit\`a di Camerino, Dipartimento di Fisica, I-62032 Camerino, Italy }
\address {$^{122}$Southern University and A\&M College, Baton Rouge, LA 70813, USA }
\address {$^{123}$College of William and Mary, Williamsburg, VA 23187, USA }
\address {$^{124}$Instituto de F\'\i sica Te\'orica, University Estadual Paulista/ICTP South American Institute for Fundamental Research, S\~ao Paulo SP 01140-070, Brazil }
\address {$^{125}$University of Cambridge, Cambridge CB2 1TN, United Kingdom }
\address {$^{126}$IISER-Kolkata, Mohanpur, West Bengal 741252, India }
\address {$^{127}$Rutherford Appleton Laboratory, HSIC, Chilton, Didcot, Oxon OX11 0QX, United Kingdom }
\address {$^{128}$Whitman College, 345 Boyer Avenue, Walla Walla, WA 99362 USA }
\address {$^{129}$National Institute for Mathematical Sciences, Daejeon 305-390, Korea }
\address {$^{130}$Universit\'e de Lyon, F-69361 Lyon, France }
\address {$^{131}$Hobart and William Smith Colleges, Geneva, NY 14456, USA }
\address {$^{132}$Janusz Gil Institute of Astronomy, University of Zielona G\'ora, 65-265 Zielona G\'ora, Poland }
\address {$^{133}$King's College London, University of London, London WC2R 2LS, United Kingdom }
\address {$^{134}$Andrews University, Berrien Springs, MI 49104, USA }
\address {$^{135}$Universit\`a di Siena, I-53100 Siena, Italy }
\address {$^{136}$Trinity University, San Antonio, TX 78212, USA }
\address {$^{137}$University of Washington, Seattle, WA 98195, USA }
\address {$^{138}$Kenyon College, Gambier, OH 43022, USA }
\address {$^{139}$Abilene Christian University, Abilene, TX 79699, USA }

%\end{document}

%% file: LVCacknowledgements.tex
The authors gratefully acknowledge the support of the United States
National Science Foundation (NSF) for the construction and operation of the
LIGO Laboratory and Advanced LIGO as well as the Science and Technology Facilities Council (STFC) of the
United Kingdom, the Max-Planck-Society (MPS), and the State of
Niedersachsen/Germany for support of the construction of Advanced LIGO 
and construction and operation of the GEO600 detector. 
Additional support for Advanced LIGO was provided by the Australian Research Council.
The authors gratefully acknowledge the Italian Istituto Nazionale di Fisica Nucleare (INFN),  
the French Centre National de la Recherche Scientifique (CNRS) and
the Foundation for Fundamental Research on Matter supported by the Netherlands Organisation for Scientific Research, 
for the construction and operation of the Virgo detector
and the creation and support  of the EGO consortium. 
The authors also gratefully acknowledge research support from these agencies as well as by 
the Council of Scientific and Industrial Research of India, 
the Department of Science and Technology, India,
the Science \& Engineering Research Board (SERB), India,
the Ministry of Human Resource Development, India,
the Spanish  Agencia Estatal de Investigaci\'on,
the  Vicepresid\`encia i Conselleria d'Innovaci\'o, Recerca i Turisme and the Conselleria d'Educaci\'o i Universitat del Govern de les Illes Balears,
the Conselleria d'Educaci\'o, Investigaci\'o, Cultura i Esport de la Generalitat Valenciana,
the National Science Centre of Poland,
the Swiss National Science Foundation (SNSF),
the Russian Foundation for Basic Research, 
the Russian Science Foundation,
the European Commission,
the European Regional Development Funds (ERDF),
the Royal Society, 
the Scottish Funding Council, 
the Scottish Universities Physics Alliance, 
the Hungarian Scientific Research Fund (OTKA),
the Lyon Institute of Origins (LIO),
the National Research, Development and Innovation Office Hungary (NKFI), 
the National Research Foundation of Korea,
Industry Canada and the Province of Ontario through the Ministry of Economic Development and Innovation, 
the Natural Science and Engineering Research Council Canada,
the Canadian Institute for Advanced Research,
the Brazilian Ministry of Science, Technology, Innovations, and Communications,
the International Center for Theoretical Physics South American Institute for Fundamental Research (ICTP-SAIFR), 
the Research Grants Council of Hong Kong,
the National Natural Science Foundation of China (NSFC),
the Leverhulme Trust, 
the Research Corporation, 
the Ministry of Science and Technology (MOST), Taiwan
and
the Kavli Foundation.
The authors gratefully acknowledge the support of the NSF, STFC, MPS, INFN, CNRS and the
State of Niedersachsen/Germany for provision of computational resources.